\documentclass[aps,prd,twocolumn,nofootinbib,superscriptaddress,floatfix,10pt]{revtex4-2}

\usepackage{amsmath,amssymb,bm}
\usepackage{graphicx}
\usepackage{tikz}
\usetikzlibrary{positioning,arrows.meta}
\usepackage{booktabs}
\usepackage{url}
\usepackage{capt-of}
\usepackage{silence}
\usepackage[colorlinks=true,allcolors=blue]{hyperref}

\graphicspath{{figures/}}

\newcommand{\LCDM}{\Lambda\mathrm{CDM}}
\newcommand{\LambdaCDM}{\Lambda\mathrm{CDM}}

\newcommand{\Msun}{\ensuremath{M_\odot}}
\newcommand{\dd}{\mathop{}\!\mathrm{d}}
\newcommand{\tauddm}{\ensuremath{\tau_{\mathrm{ddm}}}}
\newcommand{\posterior}[3]{\ensuremath{#1^{+#3}_{-#2}}}

\begin{document}

\title{Spherical Collapse and Halo Formation in a Cosmology with Decaying Dark Matter and a Semi-Cosmographic Dark Energy}
\author{Mohit Yadav}
\email{p20210462@pilani.bits-pilani.ac.in}
\affiliation{Department of Physics, Birla Institute of Technology and Science, Pilani, Rajasthan 333031, India}
\author{Tapomoy Guha Sarkar}
\email{tapomoy1@pilani.bits-pilani.ac.in}
\affiliation{Department of Physics, Birla Institute of Technology and Science, Pilani, Rajasthan 333031, India}

\begin{abstract}
We investigate nonlinear structure formation in a cosmological model combining one-body decaying dark matter (DDM) with a semi-cosmographic reconstruction of dark energy. In this scenario, a nonrelativistic dark-matter component decays into relativistic dark radiation with decay rate $\Gamma=\tau_{\rm ddm}^{-1}$, while the dark-energy sector is reconstructed directly from the expansion history rather than being fixed to a cosmological constant. Using DESI DR1 BAO and compressed ShapeFit measurements, we constrain the background evolution and propagate the resulting posterior into the nonlinear regime through spherical collapse and halo abundance calculations. This provides a unified framework connecting a reconstructed dark-energy sector and decaying dark matter (DDM) to the nonlinear formation of cosmic structures. We find that the reconstructed dark-energy equation of state can deviate from the 
$\Lambda$CDM value, $w=-1$, while the critical density threshold for collapse remains close to its standard prediction. The most pronounced signatures emerge in the abundance of massive halos, reflecting modifications to the growth of structure driven by both dark-matter decay and dynamical dark energy. By combining DESI DR1 clustering constraints with halo mass function measurements from the DESI Legacy Imaging Surveys DR9, we obtain joint constraints on the DDM lifetime and dark-energy parameters, demonstrating that halo abundances provide a powerful complementary probe of non-standard dark-sector physics.

\end{abstract}

\maketitle

\section{Introduction}
\label{sec:intro}
The formation of virialized dark-matter halos in the nonlinear regime of cosmological structure formation depends on the background expansion history, the growth of matter perturbations, and the dynamics of gravitational collapse. Consequently, the halo mass distribution provides a sensitive probe of dark-sector physics beyond stable cold dark matter and of the processes that shape cosmic structure formation. Halo formation also affects the formation of the first luminous sources and hence affects the reionization history \citep{Barkana_2001, Choudhury_2022}.

 Cold dark matter (CDM) is known to play the dominant role in the standard cosmological model of structure formation. However,  its nature is still unknown, and several observational tensions have motivated the study of non-standard dark matter scenarios \citep{2005PhR...405..279B,2015PNAS..11212249W,2017ARA&A..55..343B,2000PhRvL..84.3760S,2000PhRvL..85.1158H}.  Among these possibilities, decaying dark matter (DDM) provides a simple extension of CDM by relaxing the assumption of exact stability \cite{Blackadder2014,vattis2019dark,vattis2019late,Vattis}. If dark matter decays on cosmological timescales, the matter density, the expansion rate, and the growth history are all modified, which can suppress structure formation \citep{2001ApJ...546L..77C,2008PhRvD..77f3514B,2010PhRvD..81j3501P}.  This is phenomenologically interesting because a reduced late-time clustering amplitude may contribute to alleviating the $S_8$ tension \citep{Hildebrandt2017,Asgari2021,Heymans2021,Abdalla2022}, while changes in the expansion history can also affect inferences related to the Hubble tension \citep{Riess2016,DiValentino2021,Verde2019,Hildebrandt2017,Asgari2021,Abdalla2022,Heymans2021}.

Most earlier works \cite{Oguri2003}, which aim to constrain the decaying dark matter properties, typically use the dark energy sector to either be the cosmological constant or a fluid with some equation of state characterized by model parameters \citep{wang,pw}. 
In the present work, we do not {\it a priori} assume any dark energy equation of state.
We adopt a partially model-agnostic approach \cite{pankaj1,pankaj2} and  propagate a semi-cosmographically reconstructed dark-energy sector and a  DDM model to reconstruct spherical-collapse and halo-abundance observables from data.  We do not fix the late-time accelerated component to a cosmological constant.  The matter sector is assumed to be a one-body DDM, while the smooth dark-energy sector is reconstructed from a Pad\'e kinematic expansion for the angular diameter distance.  After subtracting the contributions of baryons, the DDM mother component, and the decay-radiation daughter from the corresponding Hubble expansion history,  the remaining smooth residual density is assigned to the effective dark-energy component \cite{pankaj1}.  The reconstructed expansion history is then carried into the nonlinear collapse regime through $\delta_c(z)$ and $\Delta_{\rm vir}(z)$, and finally the linear power spectrum is used to compute the final Halo Mass Function (HMF)  response \cite{PressSchechter1974,Despali2016} .

The spherical-collapse model gives a useful bridge between the background cosmology and the nonlinear halo population.  In its standard form, it follows the evolution of a spherical overdense region until turnaround, collapse, and virialization~\citep{1972ApJ...176....1G,1998ApJ...495...80B}.  The resulting critical collapse threshold $\delta_c$ and virial overdensity $\Delta_{\rm vir}$ enter halo abundance calculations and therefore provide a compact way to track how a non-standard background changes nonlinear structure formation. Earlier works have shown that for one-body decaying dark matter (DDM), the decay affects both the expansion history and cluster abundances~\citep{Oguri2003, Pace2020}.

In this work, we constrain a semi-cosmographic dark energy model in the presence of decaying dark matter (DDM) using a combination of DESI DR1 clustering data and halo abundance measurements from the DESI Legacy Imaging Surveys DR9. The DDM decay lifetime $\tau$ and the dark energy parameters modify the expansion history $H(z)$, which affects the linear matter power spectrum $P(k,z)$, the mass variance $\sigma(M,z)$, and the spherical-collapse threshold $\delta_c(z)$. These quantities determine the peak height, $\nu=\delta_c/\sigma$, and the resulting halo mass function, which we model using the Despali prescription.

We use DESI BAO measurements \cite{DESIFS,DESILYA,DESIFScons}  of $D_M/r_d$, $D_H/r_d$, and $D_V/r_d$, and the compressed ShapeFit data providing complementary information on structure growth and the shape of the matter power spectrum through $f\sigma_8$ and $m+n$ \citep{DESIBAO,DESILYA,Brieden2021,DESIFS,DESIFScons}. For each parameter set, we compute $P(k,z)$ with CLASS \cite{Lesgourgues2011,Blas2011}, evaluate $\sigma(M,z)$, determine $\delta_c(z)$ from spherical collapse, and construct the halo mass function. The combined DESI BAO  distances, clustering and halo abundance data are then used to place joint constraints on the dark energy and DDM parameters.

The paper is organized as follows.  Section~\ref{sec:model} gives the one-body DDM background and the semi-cosmographic reconstruction.  Section~\ref{sec:data} describes the DESI DR1 data and ShapeFit growth observables.  Section~\ref{sec:collapse} gives the spherical-collapse and virialization prescription.  Section~\ref{sec:hmf} describes the halo mass-function calculation.  Section~\ref{sec:results} presents the main results, followed by a dicussion in Sec.~\ref{sec:discussion} and finally Sec.~\ref{sec:conclusion} summarizes the conclusions.

\section{Formalism: semi-cosmographic framework with one-body DDM and dynamical dark energy}
\label{sec:model}

\subsection{One-body decaying dark matter}
\label{subsec:ddm_background}
The possibility of decaying dark matter is motivated both by fundamental particle physics and by current cosmological observations. From a theoretical perspective, absolute stability generally requires an exact protecting symmetry, whereas metastable dark matter naturally arises in many extensions of the Standard Model, often with lifetimes much longer than the age of the Universe \citep{ibarra2013indirect, Abell, clark2021cosmological}. From an observational perspective, persistent tensions within the standard $\Lambda$CDM framework, particularly those related to the expansion history and the growth of cosmic structures, have motivated investigations of departures from perfectly stable cold dark matter. A slowly decaying dark-matter component can modify the late-time matter density, expansion rate, and structure-growth history, thereby potentially alleviating some of these tensions while remaining consistent with existing observational constraints \citep{ALONSO,GNEDIN,CROFT,mohit}. Consequently, decaying dark matter provides a physically motivated and observationally testable extension of $\Lambda$CDM, making it a natural sector in which to incorporate prior physical insight while reconstructing the less understood dark-energy component.

We model the dark-matter sector as a one-body decaying dark-matter scenario.  The unstable nonrelativistic mother component, denoted by `$m$', decays into a relativistic daughter component, denoted by `${\rm dr}$', with constant decay rate $\Gamma\equiv \tauddm^{-1}$.  At the homogeneous level the two components obey \cite{cen,Oguri2003}
\begin{align}
\dot{\rho}_m+3H\rho_m&=-\Gamma\rho_m,\label{eq:ddm_mother_cont}\\
\dot{\rho}_{\rm dr}+4H\rho_{\rm dr}&=\Gamma\rho_m.\label{eq:ddm_dr_cont}
\end{align}
These equations have solutions
\begin{equation}
\rho_m(a,t)=\rho_{m0}a^{-3}\exp[-\Gamma(t-t_0)],
\label{eq:rho_m}
\end{equation}
\begin{equation}
\rho_{\rm dr}(a,t)=\Gamma\rho_{m0}a^{-4}
\int_0^t a(t')\exp[-\Gamma(t'-t_0)]\,\dd t'.
\label{eq:rho_dr}
\end{equation}
Here,  $\rho_{m0}=\rho_{c0}\Omega_{m0}$ denotes the present surviving  density of mother particles.  
The dark-energy sector is kept smooth, unclustered, but dynamic. In this work, we do not adopt any known dark-energy equation of state $w_\phi$.  
The evolution of dark energy given by \begin{equation}
\dot{\rho}_\phi+3H(1+w_\phi)\rho_\phi=0,
\label{eq:phi_cont}
\end{equation}
so that, if 
$\rho_\phi(z)=\rho_{\phi0} f_\phi(z)$, 
we have
\begin{equation}
f_\phi(z)=\exp\left[
3\int_0^z \frac{1+w_\phi(z')}{1+z'}\,\dd z'
\right].
\label{eq:f_phi}
\end{equation}
For a spatially flat Universe containing one-body DDM, baryons, radiation, and smooth dark energy, the physical background expansion may be written as
\begin{align}
\frac{H^2(z)}{H_0^2}=&
\frac{\rho_m(z)}{\rho_{c0}}
+\frac{\rho_{\rm dr}(z)}{\rho_{c0}}\nonumber\\
&+\Omega_{b0}(1+z)^3
+\Omega_{r0}(1+z)^4\nonumber\\
&+\Omega_{\phi0}f_\phi(z).
\label{eq:E2}
\end{align}
where $\rho_{c0}=3H_0^2/(8\pi G)$.

The present-epoch flatness condition is
\begin{equation}
1=\Omega_{m0}+\Omega_{{\rm dr},0}+\Omega_{b0}+\Omega_{\phi0}+\Omega_{r0}.
\label{eq:flatness}
\end{equation}

\subsection{Semi-cosmography in the one-body DDM scenario}
\label{subsec:pade}
While the $\Lambda$CDM model remains remarkably successful in describing a wide range of cosmological observations, persistent theoretical challenges and observational tensions have motivated alternatives and reconstruction techniques. Purely data-driven approaches, such as Gaussian-process reconstructions \cite{Holsclaw_2010, Holsclaw_2011_GPR, Shafieloo_2012_GPR}, minimize theoretical assumptions, but they do not readily incorporate established physical knowledge about known cosmic components. Traditional cosmographic methods provide a kinematic description of the expansion history, but they do not by themselves separate the dark-matter and dark-energy contributions. In this work, we adopt a semi-cosmographic framework: the dark-energy component remains flexible, while the matter sector is specified by a physically motivated one-body DDM model. This reduces degeneracies and lets the data reveal possible departures from the standard cosmological paradigm with minimal bias.
The standard pure cosmographic approach describes the background evolution kinematically, by expanding observable quantities such as the luminosity distance or the Hubble expansion rate \cite{Visser2005-cosmography, Visser_2015_cosmography, Dunsby_Luongo_2016_cosmography, Capozziello_2018_cosmography}. Pad\'e cosmography replaces ordinary Taylor expansions by rational functions~\cite{Pade1892,Capozziello2019,Benetti2019}, which are better behaved over the redshift range probed by BAO measurements.  By itself, however, a kinematic expansion does not identify which part of the expansion history is due to dark matter and which part is due to dark energy.

Our objective is not complete agnosticism, but a controlled hierarchy of assumptions. We incorporate phenomenologically and particle-physics-motivated modifications in the dark-matter sector, for which both theoretical motivation and observational constraints already exist, while retaining maximal flexibility in the dark-energy sector, whose fundamental physical origin remains entirely unknown. The resulting framework exploits existing knowledge where available and minimizes assumptions where ignorance is greatest.

The semi-cosmographic construction \citep{pankaj1,pankaj2}  used here has two ingredients.  The first is a Pad\'e representation of the expansion history.  The second is the physical one-body DDM matter sector.  The difference between the Pad\'e expansion rate and the known DDM, baryon, and radiation contributions is assigned to a smooth residual dark-energy density.  Its effective equation of state is then obtained by inverting the generalized Friedmann equations. 

We introduce
\begin{equation}
\xi\equiv{1+z},
\label{eq:xi}
\end{equation}
and use the Pad\'e comoving angular-diameter distance approximation~\cite{Saini2000}
\begin{equation}
D_M^{\rm P}(z)
=
\frac{2c}{H_0}
\frac{
\xi-a_1\sqrt{\xi}-1+a_1
}
{
b_1\xi+c_1\sqrt{\xi}+2-a_1-b_1-c_1
}.
\label{eq:DM_pade_xi}
\end{equation}

The corresponding Pad\'e Hubble expansion rate is
\begin{equation}
H^{\rm P}(z)
=
c\left[
\frac{\dd D_M^{\rm P}(z)}{\dd z}
\right]^{-1}.
\label{eq:HP_def}
\end{equation}

Equivalently, defining
\begin{align}
A(z)&=b_1(1+z)+c_1\sqrt{1+z}+2-a_1-b_1-c_1,\label{eq:A}\\
B(z)&=(1+z)-a_1\sqrt{1+z}-1+a_1,\label{eq:B}
\end{align}
we have
\begin{equation}
D_M^{\rm P}(z)=\frac{2c}{H_0}\frac{B(z)}{A(z)},
\label{eq:DM}
\end{equation}
\begin{equation}
H^{\rm P}(z)=\frac{H_0}{2}
\frac{A(z)^2}{
A(z)\left(1-\frac{a_1}{2\sqrt{1+z}}\right)
-B(z)\left(b_1+\frac{c_1}{2\sqrt{1+z}}\right)}.
\label{eq:H}
\end{equation}

Thus,~$D_M^{\rm P}(z)\equiv D_M^{\rm P}(z;h,a_1,b_1,c_1)$ and $H^{\rm P}(z)\equiv H^{\rm P}(z;h,a_1,b_1,c_1)$, where $h=H_0/(100~{\rm km~s^{-1}~Mpc^{-1}})$.

The semi-cosmographic step is to replace the physical $H(z)$ in Eq.~\eqref{eq:E2} by the Pad\'e quantity $H^{\rm P}(z)$.
The residual dark-energy density is
\begin{align}
\Omega_\phi^{\rm P}(z)=&\left[E^{\rm P}(z)\right]^2
-\frac{\rho_m(z)}{\rho_{c0}}
-\frac{\rho_{\rm dr}(z)}{\rho_{c0}}\nonumber\\
&-\Omega_{b0}(1+z)^3
-\Omega_{r0}(1+z)^4.
\label{eq:Omega_phi}
\end{align}
The semi-cosmographic dark-energy equation of state is then reconstructed from the conservation equation as
\begin{equation}
w_\phi^{\rm P}(z)=-1+\frac{1+z}{3\Omega_\phi^{\rm P}(z)}
\frac{\dd\Omega_\phi^{\rm P}}{\dd z}.
\label{eq:w_phi}
\end{equation}

The reconstructed semi-cosmographic dark-energy equation of state therefore has the parameter dependence
\begin{equation}
w_\phi^{\rm P}(z)\equiv
w_\phi^{\rm P}(z;h,\tauddm,\Omega_{\rm dm,ini},a_1,b_1,c_1).
\label{eq:w_phi_params}
\end{equation}

 In the MCMC, physical samples are required to keep $\Omega_\phi^{\rm P}(z)>0$ and approximately $-2\leq w_\phi^{\rm P}(z)\leq0$ over the reconstruction range.  Cosmological observables are then computed using the Pad\'e distances and the associated semi-cosmographic DDM plus dark-energy background, and the resulting predictions are fitted to the DESI DR1 data.

\begin{figure*}[t]
\centering
\resizebox{\textwidth}{!}{%
\begin{tikzpicture}[
    node distance=0.72cm,
    smallbox/.style={draw, rounded corners=2pt, thick, minimum width=4.6cm, text width=5.2cm, minimum height=0.85cm, align=center, inner sep=5pt},
    box/.style={draw, rounded corners=2pt, thick, minimum width=5.9cm, text width=8.3cm, minimum height=0.95cm, align=center, inner sep=6pt},
    bigbox/.style={draw, rounded corners=2pt, thick, minimum width=6.0cm, text width=8.4cm, minimum height=3.0cm, align=center, inner sep=7pt},
    arrow/.style={-{Latex[length=3.5mm, width=2.3mm]}, very thick}
]

\node (p0) [box, fill=blue!8] {\textbf{One-body DDM scenario with dynamical DE}};

\node (p1) [bigbox, below=of p0, fill=blue!2] {Hubble expansion rate is\\[0.6ex]
$\displaystyle
\begin{aligned}
H(z)=H_0\Biggl[&\frac{\rho_m(z)}{\rho_{c0}}
+\frac{\rho_{\rm dr}(z)}{\rho_{c0}}
+\Omega_{b0}(1+z)^3\\
&+\Omega_{r0}(1+z)^4+\Omega_\phi(z)
\Biggr]^{1/2}
\end{aligned}$\\[0.8ex]
$H(z)\equiv H(z;h,\tauddm,\Omega_{\rm dm,ini})$, with $\Gamma=\tauddm^{-1}$\\[0.8ex]
Smooth dark-energy density and EoS\\[0.5ex]
$\displaystyle
\begin{aligned}
\Omega_\phi(z)=\frac{H^2(z)}{H_0^2}-\Biggl[&\frac{\rho_m(z)}{\rho_{c0}}
+\frac{\rho_{\rm dr}(z)}{\rho_{c0}}\\
&+\Omega_{b0}(1+z)^3+\Omega_{r0}(1+z)^4
\Biggr]
\end{aligned}$\\[0.8ex]
$\displaystyle
w_\phi(z)=-1+\frac{1+z}{3\Omega_\phi(z)}\frac{\dd\Omega_\phi}{\dd z}$};

\node (p2) [box, below=of p1, fill=blue!2] {Semi-Cosmography $\equiv$ keep the one-body DDM sector fixed by $(h,\tauddm,\Omega_{\rm dm,ini})$ and replace $H(z)$ by $H^{\mathcal P}(z)$\\[0.5ex]
Dark matter: one-body DDM mother $+$ relativistic daughter\\
Dark energy: no parametric model assumed};

\node (r1) [box, right=1.6cm of p0, fill=green!12] {\textbf{Cosmography}};

\node (r2) [box, below=of r1, fill=green!3] {$\displaystyle
\xi=1+z$\\[0.5ex]
$\displaystyle
D_M^{\mathcal P}(z)=\frac{2c}{H_0}
\frac{\xi-a_1\sqrt{\xi}-1+a_1}
{b_1\xi+c_1\sqrt{\xi}+2-a_1-b_1-c_1}$\\[0.8ex]
$D_M^{\mathcal P}(z)\equiv D_M^{\mathcal P}(z;h,a_1,b_1,c_1)$\\[0.8ex]
$\displaystyle
H^{\mathcal P}(z)=c\left[\frac{\dd D_M^{\mathcal P}(z)}{\dd z}\right]^{-1}$\\[0.8ex]
$H^{\mathcal P}(z)\equiv H^{\mathcal P}(z;h,a_1,b_1,c_1)$};

\node (r3) [smallbox, below=of r2, fill=orange!14] {\textbf{Semi-Cosmography}};

\node (r4) [box, below=of r3, fill=orange!3] {$\displaystyle
w_\phi^{\mathcal P}(z)=
-1+\frac{1+z}{3\Omega_\phi^{\mathcal P}(z)}
\frac{\dd\Omega_\phi^{\mathcal P}}{\dd z}$\\[0.8ex]
where\\[0.5ex]
$\displaystyle
\begin{aligned}
\Omega_\phi^{\mathcal P}(z)=&\left[\frac{H^{\mathcal P}(z)}{H_0}\right]^2
-\frac{\rho_m(z)}{\rho_{c0}}
-\frac{\rho_{\rm dr}(z)}{\rho_{c0}}\\
&-\Omega_{b0}(1+z)^3-\Omega_{r0}(1+z)^4
\end{aligned}$\\[0.8ex]
$w_\phi^{\mathcal P}(z)\equiv
w_\phi^{\mathcal P}(z;h,\tauddm,\Omega_{\rm dm,ini},a_1,b_1,c_1)$};

\node (r5) [box, below=of r4, fill=orange!3] {Calculate observables using the one-body DDM densities and the semi-cosmographic dark-energy EoS\\[0.5ex]
Use the CLASS code to calculate $P(k,z)$ and thus
$w_\phi^{\mathcal P}(z)\rightarrow
\{D_M^{\mathcal P}/r_d,\,D_H^{\mathcal P}/r_d,\,D_V^{\mathcal P}/r_d,\,D_H^{\mathcal P}/D_M^{\mathcal P},\,f\sigma_{\rm s8},\,m+n\}$};

\node (r6) [box, below=of r5, fill=orange!3] {Fit the DESI DR1 BAO and DESI DR1 ShapeFit observables\\[0.5ex]
and constrain the one-body DDM, growth-amplitude, and Pad\'e parameters\\[0.5ex]
$\left(\log_{10}(\tauddm/{\rm yr}),h,\Omega_{\rm dm,ini},\sigma_{8,0},a_1,b_1,c_1\right)$ using MCMC \\
$\mathrm{CLASS}
\longrightarrow  \sigma(M,z), \delta_c(z), \Delta_{vir}
\longrightarrow \mathrm{HMF}$\\ Halo abundance data joint constraints.
};

\draw[arrow] (p0) -- (p1);
\draw[arrow] (p1) -- (p2);
\draw[arrow] (r1) -- (r2);
\draw[arrow] (r2) -- (r3);
\draw[arrow] (r3) -- (r4);
\draw[arrow] (r4) -- (r5);
\draw[arrow] (r5) -- (r6);
\draw[arrow] (p2.east) -- ++(0.75,0) |- (r4.west);

\end{tikzpicture}%
}
\caption{Schematic flowchart of the semi-cosmographic method used in this work.  The expansion history is described directly by the Pad\'e comoving angular-diameter distance and the associated $H^{\mathcal P}(z)$, while the matter sector is fixed to the one-body DDM model.  The smooth dark-energy equation of state is reconstructed as the residual component required by the Pad\'e expansion history after subtracting the one-body DDM mother density, the relativistic decay-radiation density, baryons, and standard radiation.}
\label{fig:Flowchart}
\end{figure*}
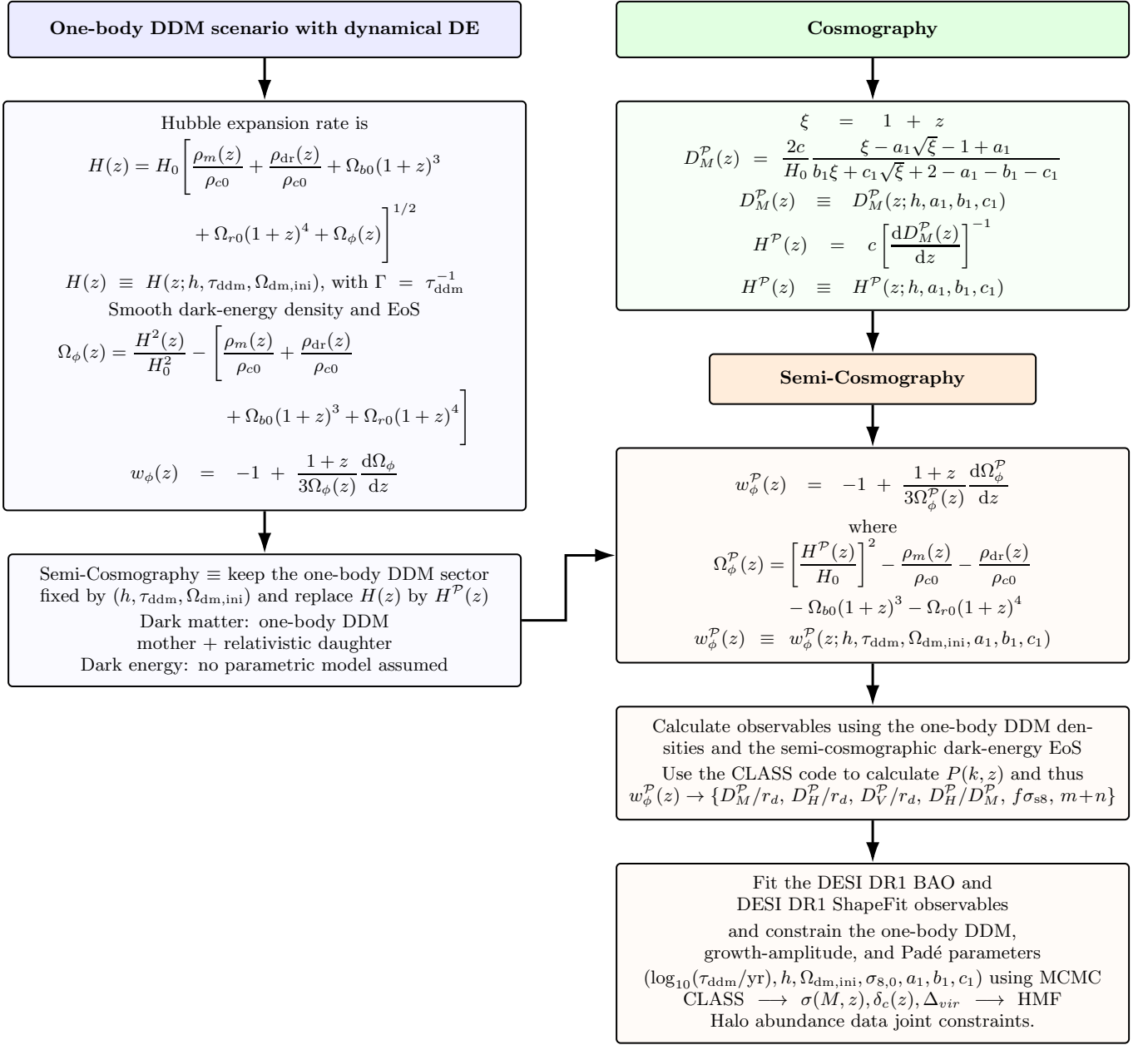

The full one-body semi-cosmographic pipeline is summarized in Fig.~\ref{fig:Flowchart}.

A natural question is why one should adopt a semi-cosmographic framework rather than a completely model-agnostic reconstruction of both the dark-energy and dark-matter sectors. Our choice is motivated by several considerations.
\begin{itemize}
\item  The dark-energy and dark-matter sectors are not on the same epistemic footing. While the existence of dark matter is firmly established through multiple independent observations, the physical origin and dynamics of dark energy remain largely unknown. It is therefore reasonable to remain maximally agnostic about dark energy while incorporating physically motivated information about the matter sector.

\item Decaying dark matter is not an ad hoc assumption but a phenomenologically and theoretically motivated extension of the standard cosmological model. Many particle-physics scenarios naturally allow extremely long-lived dark matter particles whose decay timescales are comparable to or larger than the age of the Universe.

\item Unlike dark energy, the decaying dark matter sector is already subject to direct observational constraints from the CMB, large-scale structure, weak lensing and galaxy-clustering measurements. Consequently, this sector possesses a degree of empirical guidance that can be incorporated into the reconstruction.

\item Complete model independence is not equivalent to complete absence of assumptions. Every reconstruction method necessarily contains implicit assumptions regarding the background framework, parameterization, kernels, priors or smoothness conditions. The relevant question is therefore not whether assumptions are present, but whether they are physically motivated and minimized.

\item Simultaneously reconstructing both dark-energy and dark-matter sectors can introduce severe degeneracies, making it difficult to identify which component is responsible for a departure from $\Lambda$CDM. Incorporating a phenomenologically motivated matter sector helps break these degeneracies and improves the interpretability of the reconstruction.

\item Our objective is to probe the least understood component of the cosmic energy budget. By constraining the matter sector using physically motivated assumptions, the available freedom is concentrated in the dark-energy sector, allowing the data to reveal possible departures from a cosmological constant more effectively.

\item The adopted framework also serves as a diagnostic tool. Rather than asking whether a particular dark-energy model fits the data, we investigate the converse question: given a phenomenologically motivated decaying dark matter sector, what form of dark-energy behavior is preferred by the observations? This allows the data to indicate whether $\Lambda$ remains sufficient or whether non-trivial dark-energy dynamics is required.

\item The semi-cosmographic approach therefore occupies a useful middle ground between fully parametric cosmological models and completely non-parametric reconstructions. It preserves the exploratory power of model-independent methods while retaining well-motivated physical information from sectors of cosmology that are already comparatively well understood.
\end{itemize}

\section{Galaxy Data}
\label{sec:data}

To constrain the one-body DDM plus semi-cosmographic dark-energy background, we use the compressed clustering measurements from DESI Data Release 1.  The data set has two parts.  Firstly, we use the DESI DR1 BAO distance measurements from galaxies, quasars, and the Lyman-$\alpha$ forest~\cite{DESIBAO,DESILYA}.  These measurements constrain the background geometry through ratios of cosmological distances to the sound horizon at the baryon-drag epoch.  Secondly, we use the DESI DR1 ShapeFit compression of the full-shape clustering information~\cite{Brieden2021,DESIFS,DESIFScons}.  From the ShapeFit vector, we keep only the growth and shape observables, $f\sigma_{\rm s8}$ and $m+n$, while the distance-like ShapeFit entries are not included in order to avoid double-counting the geometric information already supplied by the BAO data block.

Operationally, each MCMC sample defines a Pad\'e expansion history and a one-body DDM matter sector.  For that sample we compute the BAO distance ratios, solve the linear growth equation for $f\sigma_{\rm s8}$, and evaluate the ShapeFit shape combination $m+n$ using the same fixed $n_s$ and baryon density used elsewhere in the analysis. The power spectrum is computed using CLASS.  The BAO and ShapeFit entries are then compared with the DESI compressed data vector using the corresponding covariance submatrix.  The DESI data, therefore, enter only through the background, growth, and broadband-shape constraints; the cluster abundance curves shown later are predictions obtained after propagating the DESI-constrained posterior into spherical collapse and the HMF.
Further, we also use data on the Halo Mass Function jointly with the BAO + ShapeFit data. For this analysis the MCMC requires the evaluation of the HMF also for each sample. 
\subsection{BAO distances}

The BAO observables used in this work are distances $D_M$, $D_H$ and $D_V$.   For a spatially flat background,
\begin{equation}
D_M(z)=c\int_0^z\frac{\dd z'}{H(z')},
\label{eq:DM_data}
\end{equation}
while the radial Hubble distance and the volume-averaged distance are given by 
\begin{equation}
D_H(z)=\frac{c}{H(z)},
\label{eq:DH}
\end{equation}
\begin{equation}
D_V(z)=\left[zD_M(z)^2D_H(z)\right]^{1/3},
\label{eq:DV}
\end{equation}
respectively.  The theoretical predictions for $D_V/r_d$, $D_M/r_d$, and $D_H/r_d$ are compared with the DESI compressed vector.  The sound horizon is fixed to the fiducial value $ r_d=147.089\;{\rm Mpc}.$
This fixed value is a phenomenological standard-ruler calibration for the compressed DESI likelihood, not a self-consistent early-Universe prediction recomputed for every DDM sample.  The first redshift bin and the QSO bin enter as isotropic measurements through $D_V/r_d$, while the LRG, LRG+ELG, ELG, and Lyman-$\alpha$ bins enter through the anisotropic pair $(D_M/r_d,D_H/r_d)$. The BAO data vector and its covariance are taken from the DESI DR1 BAO~\citep{DESIBAO,DESILYA,DESIBAOCosmo}. 
For the ShapeFit sector, we use the compressed DESI DR1 ShapeFit data vector and the corresponding Gaussian covariance reported in Appendix A of the DESI DR1 full-shape analysis~\citep{Brieden2021,DESIFS,DESIFScons}.

\subsection{ShapeFit growth and shape information}

The DESI full-shape analysis compresses the broadband clustering information into a small number of parameters in each redshift bin.  In the ShapeFit basis, the compressed vector contains distance information, the redshift-space growth amplitude $f\sigma_{\rm s8}$, and the shape parameter combination $m+n$.  Since the BAO data block already carries the distance information used in this analysis, we use only the 12 ShapeFit entries on  $f\sigma_{\rm s8}$ and $m+n$ for the BGS, LRG, ELG, and QSO samples \citep{Brieden2021,DESIFS,DESIFScons}.

For the growth observable, we solve the standard linear equation for the growing mode $D_+(a)$.  Primes denote derivatives with respect to $\ln a$:
\begin{equation}
D_+''+\left(2+\frac{\dd\ln H}{\dd\ln a}\right)D_+'
-\frac{3}{2}\Omega_{\rm m}^{\rm cl}(a)D_+=0 .
\label{eq:growth_D}
\end{equation}
where, 
\begin{equation}
\Omega_{\rm m}^{\rm cl}(a)=
\frac{\Omega_{b0} a^{-3}+\rho_m(a)/\rho_{c0}}{E^2(a)}, ~~{\rm with} ~~\rho_{c0} = \frac{3H_0^2}{8\pi G}.
\label{eq:Omega_m_cl}
\end{equation}
The crucial observable entering ShapeFit is
\begin{align}
f\sigma_{\rm s8}(z)&=f(z)\sigma_{\rm s8}(z),\label{eq:fs8}\\
{\rm where} ~~f(z)&=\frac{\dd\ln D_+}{\dd\ln a}, ~~{\&}~~~
\sigma_{\rm s8}(z)=D_+(z)\sigma_{{\rm s8},0}.\nonumber
\end{align}
The second ShapeFit observable used here, $m+n$, captures the broadband power-spectrum-shape information beyond the BAO peak position and the RSD amplitude.  In the numerical analysis we compute this quantity from the logarithmic slope of the Eisenstein-Hu no-wiggle transfer function at the ShapeFit pivot scale, with the same fixed $n_s$ used throughout the MCMC.  This makes the ShapeFit block complementary to the BAO block: BAO constrains the geometric expansion history, while $f\sigma_{\rm s8}$ and $m+n$ add growth and shape information.  The final posterior therefore uses 24 compressed DESI DR1 data points: 12 BAO distance measurements and 12 ShapeFit growth/shape measurements.

\subsection{Halo Mass Function Data}
The halo mass function data used in this work are taken from Halo Properties and Mass Functions of Groups/Clusters from the DESI Legacy Imaging Surveys DR9 \citep{DR9}. This study constructs one of the largest homogeneous catalogs of galaxy groups and clusters identified in the DESI Legacy Imaging Surveys DR9, spanning a wide halo mass range. Using calibrated halo mass estimates, it provides measurements of the observed halo mass function over a broad range of masses and redshifts, making it a valuable dataset for testing cosmological models through comparisons with theoretical halo abundance predictions. From this work, we use the observed halo mass function (HMF) measurements, namely the binned halo number densities as a function of halo mass, together with their corresponding uncertainties. These measurements are compared with our theoretical halo mass function predictions to constrain the DDM and the dark energy sectors.
\section{Spherical collapse and virialization}
\label{sec:collapse}

The spherical-collapse calculation supplies the two nonlinear inputs used by the Halo Mass Function (HMF): the critical linear threshold $\delta_c(z)$ and the virial overdensity $\Delta_{\rm vir}(z)$.  We follow the standard top-hat construction, adapted to the one-body DDM with a  semi-cosmographic smooth dark-energy background defined in Sec.~\ref{sec:model}~\citep{1972ApJ...176....1G,1998ApJ...495...80B,Oguri2003,Pace2020}.  

The Universe comprises the surviving nonrelativistic DDM mother with density $\rho_m$, the relativistic daughter with density $\rho_{\rm dr}$, baryons with density $\rho_b$, and a smooth effective dark-energy component with density $
\rho_\phi(z)=\rho_{\phi0}f_\phi(z)$ 
where $f_\phi(z)$, defined in Eq.~\eqref{eq:f_phi}, encapsulates the dynamical evolution of dark energy.  

The background density of the components that cluster is then given by
\begin{equation}
\bar{\rho}_{\rm cl}(a)= \bar\rho_m(a)+\bar \rho_b(a),
\label{eq:rho_cl_bar}
\end{equation}
Assuming that the clustering is of the pressureless component, the daughter radiation and the smooth dark-energy component are taken to be homogeneous on the top-hat scale.  They affect $H(a)$ and the acceleration of a spherical shell, but they do not contribute to clustering inside the top-hat.  We also note here that the dark radiation is the only radiation component in our analysis.

Let $R(t)$ be the physical radius of a spherical top-hat shell and let $M_{\rm cl}(t)$ be the total clustered mass enclosed by that shell.  The baryonic part of $M_{\rm cl}$ is conserved, while the DDM mother part decays with rate $\Gamma=\tauddm^{-1}$.  The shell acceleration is
\begin{align}
\frac{\ddot R}{R}=&
-\frac{G M_{\rm cl}(t)}{R^3}
-H_0^2\Gamma\Omega_{m0}a^{-4}{\cal I}(t)\nonumber\\
&-\frac{H_0^2}{2}\Omega_{\phi0}f_\phi(z)
\left[1+3w_\phi(z)\right].
\label{eq:R_acceleration}
\end{align}
Here $\Omega_{m0}=\rho_{m0}/\rho_{c0}$ is the present surviving-mother density parameter, $\Omega_{\phi0}=\rho_{\phi0}/\rho_{c0}$, and
\begin{equation}
{\cal I}(t)=\int_0^t a(t')\exp[-\Gamma(t'-t_0)]\,\dd t'
\label{eq:I_collapse}
\end{equation}
is the time integral that appears in the homogeneous daughter-radiation density.  The three acceleration terms in Eq.~\eqref{eq:R_acceleration} correspond, respectively, to the self-gravity of the clustered mass, the homogeneous daughter-radiation, and the active gravitational density $\rho_\phi+3p_\phi$ of the smooth dark-energy sector.

To calculate the density threshold $\delta_c$, we first look at the nonlinear density contrast $\delta$.  Consider the real top-hat sphere of radius $R$ and a hypothetical background sphere containing the same clustered material but expanding with the background.  The latter has physical radius $R_{\rm bg}\propto a$.  Since the mass $M$ is the same in both spheres the nonlinear density contrast, can be obtained from the ratio of the density of the clustering component in the top hat (`th' superscript is used)  to the density of the background sphere as 
\begin{equation}
1+\delta(a)=\frac{\rho_{\rm cl}^{\rm th}(a)}{\bar{\rho}_{\rm cl}(a)},
\quad {\rm with}\quad 
R(a)\propto a[1+\delta(a)]^{-1/3}.
\label{eq:delta_radius_relation}
\end{equation}
Differentiating the radius-density relation twice and subtracting the homogeneous background acceleration from the shell acceleration gives a closed equation for $\delta(a)$.  The homogeneous daughter radiation and smooth dark energy enter this equation through $H(a)$, while the direct gravitational source is the clustered pressureless density:
\begin{equation}
\delta''+\left(2+\frac{\dd\ln H}{\dd\ln a}\right)\delta'
-\frac{4}{3}\frac{(\delta')^2}{1+\delta}
-\frac{3}{2}\Omega_{\rm m}^{\rm cl}(a)\delta(1+\delta)=0 ,
\label{eq:deltaNL}
\end{equation}
where $' \equiv d /\dd\ln a$.  The associated linear equation is obtained by taking $\delta\ll1$, dropping the quadratic derivative term, and replacing the nonlinear contrast by the linear contrast $\delta_L$:
\begin{equation}
\delta_L''+\left(2+\frac{\dd\ln H}{\dd\ln a}\right)\delta_L'
-\frac{3}{2}\Omega_{\rm m}^{\rm cl}(a)\delta_L=0 .
\label{eq:deltaL}
\end{equation}
\begin{equation}
{\rm 
where, } ~~~~~~\Omega_{\rm m}^{\rm cl}(a)
=\frac{\bar{\rho}_{\rm cl}(a)}{\rho_{c0}E^2(a)}.
 \label{eq:Omega_cl_collapse}
\end{equation}
To compare the nonlinear collapse with its linear extrapolation, Eqs.~\eqref{eq:deltaNL} and~\eqref{eq:deltaL} are evolved from the same early perturbation.  At an early scale factor $a_i$, the amplitude is chosen to be small, $\delta_i\ll1$, so that
\begin{equation}
\delta_L(a_i)=\delta(a_i)=\delta_i .
\label{eq:collapse_initial_conditions}
\end{equation}
The initial derivatives are set by the growing mode solution of the linear evolution Eq.~\eqref{eq:deltaL}.   For a chosen collapse redshift $z_{\rm coll}$, the initial amplitude $\delta_i$ is chosen so that the nonlinear top-hat reaches formal collapse at $z_{\rm coll}$.   In the pressureless top-hat idealization this corresponds to $R\rightarrow0$, or equivalently $\delta\rightarrow\infty$.  The linear perturbation with the same initial amplitude is evolved only up to the same epoch; with its value there defines the critical collapse threshold,
\begin{equation}
\delta_c(z_{\rm coll})
=\delta_L(z_{\rm coll}).
\label{eq:deltac}
\end{equation}
Turnaround is the first epoch at which the physical radius stops expanding, $\dot R=0$.  Using Eq.~\eqref{eq:delta_radius_relation}, this condition becomes
\begin{equation}
\frac{\delta'}{1+\delta}-3=0 .
\label{eq:turnaround}
\end{equation}

The virial step is evaluated along the post-turnaround nonlinear trajectory.  Let $a_{\rm ta}$ and $R_{\rm ta}$ be the scale factor and radius at turnaround. We define
\begin{align}
x&=\frac{a}{a_{\rm ta}},
&
y&=\frac{R}{R_{\rm ta}},
&
\zeta&=\frac{\rho_{\rm cl}^{\rm th}(a_{\rm ta})}{\bar{\rho}_{\rm cl}(a_{\rm ta})}
=1+\delta_{\rm ta}.
\label{eq:xy_zeta}
\end{align}
The radius-density relation gives
\begin{equation}
y=x\left(\frac{\zeta}{1+\delta}\right)^{1/3}.
\label{eq:y_traj}
\end{equation}
We also introduce the turnaround density scale $B$, the dimensionless time $\theta$, and the dimensionless decay rate $\gamma$,
\begin{align}
B^2&=\frac{8\pi G}{3}\bar{\rho}_{\rm cl}(a_{\rm ta})
=H^2(a_{\rm ta})\Omega_{\rm m}^{\rm cl}(a_{\rm ta}),
\nonumber\\
\theta&=Bt,
\qquad
\gamma=\frac{\Gamma}{B}.
\label{eq:turnaround_variables}
\end{align}
At turnaround, the clustered background is split into the mother and baryon fractions
\begin{equation}
f_m=\frac{\rho_m(a_{\rm ta})}{\bar{\rho}_{\rm cl}(a_{\rm ta})},
\qquad
f_b=\frac{\rho_b(a_{\rm ta})}{\bar{\rho}_{\rm cl}(a_{\rm ta})},
\qquad
f_m+f_b=1 .
\label{eq:cluster_fractions}
\end{equation}
The remaining clustered mass at dimensionless time $\theta$, normalized to its turnaround value, is
\begin{equation}
\mu(\theta)=f_b+f_m\exp[-\gamma(\theta-\theta_{\rm ta})],
\qquad
\theta_{\rm ta}=Bt_{\rm ta},
\label{eq:mass_fraction}
\end{equation}
where $t_{\rm ta}$ is the turnaround time.  For the smooth dark-energy contribution in the virial condition, we use
\begin{align}
z(x)&=\frac{1}{a_{\rm ta}x}-1,
&
z_{\rm ta}&=\frac{1}{a_{\rm ta}}-1,
\nonumber\\
g_\phi(x)&=\frac{\rho_\phi[z(x)]}{\rho_\phi(z_{\rm ta})}
=\frac{f_\phi[z(x)]}{f_\phi(z_{\rm ta})},
&
\eta_\phi&=\frac{\rho_\phi(z_{\rm ta})}{\bar{\rho}_{\rm cl}(a_{\rm ta})}.
\label{eq:de_virial_factors}
\end{align}
 The instantaneous virial condition may be written as 
\begin{align}
\dot R^2
=&
\frac{G M_{\rm cl}(t)}{R}
-\frac{8\pi G}{3}\rho_{\rm dr}(t)R^2
\nonumber\\
&-\frac{4\pi G}{3}\rho_\phi(t)
\left[1+3w_\phi(t)\right]R^2 .
\label{eq:virial_physical}
\end{align}
This equation has units of velocity squared.  We rewrite it in turnaround variables. 
We note that  
\begin{equation}
\frac{\dot R^2}{B^2R_{\rm ta}^2}
=\left(\frac{\dd y}{\dd\theta}\right)^2 , 
\label{eq:virial_lhs}
\end{equation}
\begin{align}
\frac{1}{B^2R_{\rm ta}^2}\frac{G M_{\rm cl}(\theta)}{R}
&=
\frac{1}{2}\zeta\mu(\theta)y^{-1}.
\label{eq:virial_self_term}
\end{align}

For the daughter-radiation term, we first rewrite the density generated by the decaying mother component relative to the turnaround scale.  Starting from Eq.~\eqref{eq:rho_dr} and using $\rho_m(a_{\rm ta})=\rho_{m0}a_{\rm ta}^{-3}\exp[-\Gamma(t_{\rm ta}-t_0)]$, we have
\begin{align}
\rho_{\rm dr}(t)
&=\Gamma\rho_m(a_{\rm ta})a_{\rm ta}^3a^{-4}
\int_0^t a(t')e^{-\Gamma(t'-t_{\rm ta})}\,\dd t'
\nonumber\\
&=\Gamma\rho_m(a_{\rm ta})x^{-4}
\int_0^t x(t')e^{-\Gamma(t'-t_{\rm ta})}\,\dd t' .
\end{align}
Changing variables to $\tilde\theta=Bt'$ gives
\begin{align}
\rho_{\rm dr}(t)
&=\gamma\,\rho_m(a_{\rm ta})x^{-4}
\int_0^\theta x(\tilde\theta)
e^{-\gamma(\tilde\theta-\theta_{\rm ta})}\,\dd\tilde\theta
\nonumber\\
&=\gamma f_m\bar{\rho}_{\rm cl}(a_{\rm ta})x^{-4}{\cal J}(\theta),
\label{eq:rho_dr_turnaround}
\end{align}
where
\begin{equation}
{\cal J}(\theta)=
\int_0^\theta x(\tilde\theta)
\exp[-\gamma(\tilde\theta-\theta_{\rm ta})]\,\dd\tilde\theta .
\label{eq:J}
\end{equation}
Therefore, 
\begin{align}
\frac{1}{B^2R_{\rm ta}^2}
\left[-\frac{8\pi G}{3}\rho_{\rm dr}R^2\right]
&=
-\gamma f_m x^{-4}{\cal J}(\theta)y^2 .
\label{eq:virial_dr_term}
\end{align}

The smooth dark-energy contribution can be written in terms of the same variables.  Since
\begin{equation}
\rho_\phi[z(x)]
=\rho_\phi(z_{\rm ta})g_\phi(x)
=\eta_\phi\bar{\rho}_{\rm cl}(a_{\rm ta})g_\phi(x),
\end{equation}
the dark-energy term becomes
\begin{align}
\frac{1}{B^2R_{\rm ta}^2}
\left[-\frac{4\pi G}{3}\rho_\phi(1+3w_\phi)R^2\right]
&=
-\frac{1}{2}\eta_\phi g_\phi(x)
\nonumber\\
&\quad\times
\left[1+3w_\phi(x)\right]y^2 .
\label{eq:virial_de_term}
\end{align}

Combining Eqs.~\eqref{eq:virial_lhs}, \eqref{eq:virial_self_term}, \eqref{eq:virial_dr_term}, and \eqref{eq:virial_de_term}, the virial point is defined as the first post-turnaround root of
\begin{align}
\left(\frac{\dd y}{\dd\theta}\right)^2
=&\frac{1}{2}\zeta\,\mu(\theta)y^{-1}
-\gamma f_m y^2x^{-4}{\cal J}(\theta)\nonumber\\
&-\frac{1}{2}\eta_\phi g_\phi(x)
\left[1+3w_\phi(x)\right]y^2 .
\label{eq:virial_condition}
\end{align}
The terms on the right-hand side of Eq.~\eqref{eq:virial_condition} therefore have a direct origin in the physical closure: self-gravity of the remaining clustered mass, homogeneous daughter-radiation gravity generated by the decaying mother component, and the active gravitational density of smooth dark energy.  The derivative entering the virial condition is evaluated from the nonlinear solution using
\begin{equation}
\frac{\dd y}{\dd\theta}
=y\left[1-\frac{\delta'}{3(1+\delta)}\right]
\frac{E(a)}
{\left[\bar{\rho}_{\rm cl}(a_{\rm ta})/\rho_{c0}\right]^{1/2}} .
\label{eq:dy_dtheta}
\end{equation}
This condition treats the smooth sector through its active gravitational density and the clustered sector through the evolving bound mass. 

We obtain the root  $y_{\rm vir}=R_{\rm vir}/R_{\rm ta}$, which is used later for the virial density calculation. At a general post-turnaround time,
\begin{equation}
\rho_{\rm cl}^{\rm th}(a)
=
\frac{M_{\rm cl,ta}\mu(\theta)}
{(4\pi/3)R_{\rm ta}^3y^3}.
\label{eq:rho_th_vir_deriv}
\end{equation}
The corresponding background comparison sphere contains the same surviving clustered mass, but its radius grows as $R_{\rm bg}=R_{\rm bg,ta}x$.  Since the turnaround density ratio is $\zeta$, the background comparison volume is
\begin{equation}
V_{\rm bg}(a)
=\frac{4\pi}{3}R_{\rm bg,ta}^3x^3
=\frac{4\pi}{3}\zeta R_{\rm ta}^3x^3 .
\end{equation}
Thus
\begin{equation}
\bar{\rho}_{\rm cl}(a)
=
\frac{M_{\rm cl,ta}\mu(\theta)}
{(4\pi/3)\zeta R_{\rm ta}^3x^3}.
\label{eq:rho_bg_vir_deriv}
\end{equation}
Dividing Eq.~\eqref{eq:rho_th_vir_deriv} by Eq.~\eqref{eq:rho_bg_vir_deriv} shows explicitly that the common post-turnaround mass factor cancels:
\begin{equation}
\frac{\rho_{\rm cl}^{\rm th}(a)}{\bar{\rho}_{\rm cl}(a)}
=\zeta\frac{x^3}{y^3}.
\label{eq:Delta_xy}
\end{equation}
At virialization, this gives the virial overdensity relative to the mean clustered density,
\begin{equation}
\Delta_{\rm vir,mean}
=\zeta\left(\frac{a_{\rm vir}}{a_{\rm ta}}\right)^3
\frac{1}{y_{\rm vir}^3}.
\label{eq:Delta_mean}
\end{equation}
The corresponding overdensity relative to the critical density is
\begin{equation}
\Delta_{\rm vir,crit}
=\Delta_{\rm vir,mean}\,\Omega_{\rm m}^{\rm cl}(a_{\rm vir}).
\label{eq:Delta_crit}
\end{equation}
The nonlinear density contrast at virialization is given by $\delta_{\rm vir}=\Delta_{\rm vir,mean}-1$.

\section{Halo mass function}
\label{sec:hmf}

The halo mass function (HMF) gives the comoving number density of collapsed objects as a function of mass, providing a direct link between theoretical collapse models and observable halo abundances. It is particularly sensitive to the underlying cosmology because the abundance of the most massive halos is governed by the exponentially suppressed high-peak tail of the density fluctuation distribution. As a result, even small changes in the collapse dynamics or growth history can lead to significant variations in the predicted number of high-mass objects.
Following the strategy used in generalized-dark-matter spherical-collapse studies~\citep{Pace2020}, we keep the standard simulation-calibrated HMF template but insert the collapse and power-spectrum quantities appropriate to the present model.  The model-dependent inputs are the critical threshold $\delta_c(z)$, the virial overdensity $\Delta_{\rm vir,mean}(z)$, and the linear power spectrum $P(k,z)$. 
These, as we have seen, have direct bearing on the DDM parameters and the effective semi-cosmographic dark energy equation of state. 

The calculation follows the standard peak-height approach introduced by Press--Schechter and later extensions, the Sheth--Tormen multiplicity form, and the spherical-overdensity calibration of Despali et al.~\citep{PressSchechter1974,Bond1991,ShethTormen1999,ShethMoTormen2001,Despali2016}.

The first ingredient is the variance of the linear density field smoothed on a real-space top-hat scale.  The halo mass is assigned to baryons plus the surviving nonrelativistic DDM mother component; the daughter radiation and the smooth dark-energy component are not counted as halo mass.  A halo of mass $M$ is assigned a smoothing radius
\begin{equation}
M=\frac{4\pi}{3}R^3(M)\bar{\rho}_{{\rm cl},0},
\qquad
R(M)=\left(\frac{3M}{4\pi\bar{\rho}_{{\rm cl},0}}\right)^{1/3},
\label{eq:R}
\end{equation}
where $\bar{\rho}_{{\rm cl},0}=\rho_{c0}(\Omega_{b0}+\Omega_{m0})$ is the present mean clustered matter density.  For the top-hat window,
\begin{equation}
W(x)=3\frac{\sin x-x\cos x}{x^3}.
\label{eq:W}
\end{equation}
The mass variance is then
\begin{equation}
\sigma^2(M,z)=
\frac{1}{2\pi^2}
\int_0^\infty k^2P(k,z)W^2[kR(M)]\,\dd k,
\label{eq:sigma_k}
\end{equation}
where $P(k,z)$ is the linear matter power spectrum returned by CLASS for the one-body DDM plus semi-cosmographic Dark Energy cosmology.  
The collapse threshold enters through the peak height
\begin{equation}
\nu(M,z)=\frac{\delta_c(z)}{\sigma(M,z)} .
\label{eq:nu}
\end{equation}
The multiplicity function is written in the Sheth--Tormen form \citep{ShethMoTormen2001}
\begin{align}
\nu f(\nu)
&=A_{\rm ST}\sqrt{\frac{2a_{\rm ST}}{\pi}}\,
\nu\exp\left(-\frac{a_{\rm ST}\nu^2}{2}\right)
\nonumber\\
&\quad\times
\left[1+(a_{\rm ST}\nu^2)^{-p_{\rm ST}}\right].
\label{eq:mult}
\end{align}
The parameters $A_{\rm ST}$, $a_{\rm ST}$, and $p_{\rm ST}$ are evaluated from the Despali et al. spherical-overdensity calibration as functions of the halo-density convention.  We do not recalibrate these coefficients for the one-body DDM plus reconstructed-DE model.  Such a recalibration would require dedicated $N$-body simulations for this dark-sector model and the same halo finder and mass definition used in the HMF measurement.  Our semi-analytic calculation therefore follows the usual approach adopted in collapse-based extensions of $\LCDM$: the fitting function is kept fixed, while the model dependence enters through $\delta_c$, $\Delta_{\rm vir}$, and $P(k,z)$.

All overdensities entering the Despali calibration are expressed relative to the mean clustered density.  If $\Delta_{\rm SO,mean}(z)$ is the chosen spherical-overdensity halo definition, then
\begin{equation}
x_\Delta(z)=
\log_{10}\left[
\frac{\Delta_{\rm SO,mean}(z)}
{\Delta_{\rm vir,mean}(z)}
\right],
\label{eq:x_despali}
\end{equation}
and
\begin{align}
a_{\rm ST}(x_\Delta)
&=0.4332x_\Delta^2+0.2263x_\Delta+0.7665,
\nonumber\\
p_{\rm ST}(x_\Delta)
&=-0.1151x_\Delta^2+0.2554x_\Delta+0.2488,
\nonumber\\
A_{\rm ST}(x_\Delta)
&=-0.1362x_\Delta+0.3292 .
\label{eq:despali_coefficients}
\end{align}
For the HMF figures and for the halo-mass likelihood we use the $180m$ spherical-overdensity convention, matching the Wang et al. halo mass-function data.  In this convention the halo radius encloses an average density $180$ times the mean clustered matter density.  Therefore the threshold entering the Despali calibration is simply

\begin{equation}
\Delta_{\rm SO,mean}^{180m}=180 .
\label{eq:delta_180m_mean}
\end{equation}

The HMF is given by 
\begin{equation}
\frac{\dd n}{\dd M}
=\frac{\bar{\rho}_{{\rm cl},0}}{M^2}\nu f(\nu)
\left|\frac{\dd\ln\nu}{\dd\ln M}\right|.
\label{eq:hmf}
\end{equation}
The derivative in Eq.~\eqref{eq:hmf} measures how the peak height changes when the halo mass is changed.  The HMF is evaluated on a broad grid, $10^{11}\leq M/(\Msun/h)\leq10^{16}$, at redshifts $z=0,0.5,1,1.5,2$ for the plotted posterior predictions.  The comparison figures focus on $10^{11}\leq M/(\Msun/h)\leq10^{15}$, where the fitted $\LambdaCDM$ comparison is stable over all redshift slices.  The HMF uses CLASS spectra~\citep{Lesgourgues2011,Blas2011} for the combined one-body DDM plus reconstructed smooth-dark-energy background.  The same $180m$ mass definition is used for the DESI-only prediction, the joint DESI plus halo-mass likelihood, and the fitted-$\LambdaCDM$ comparison.

\begin{figure*}
\begin{center}
\includegraphics[width=0.85\columnwidth]{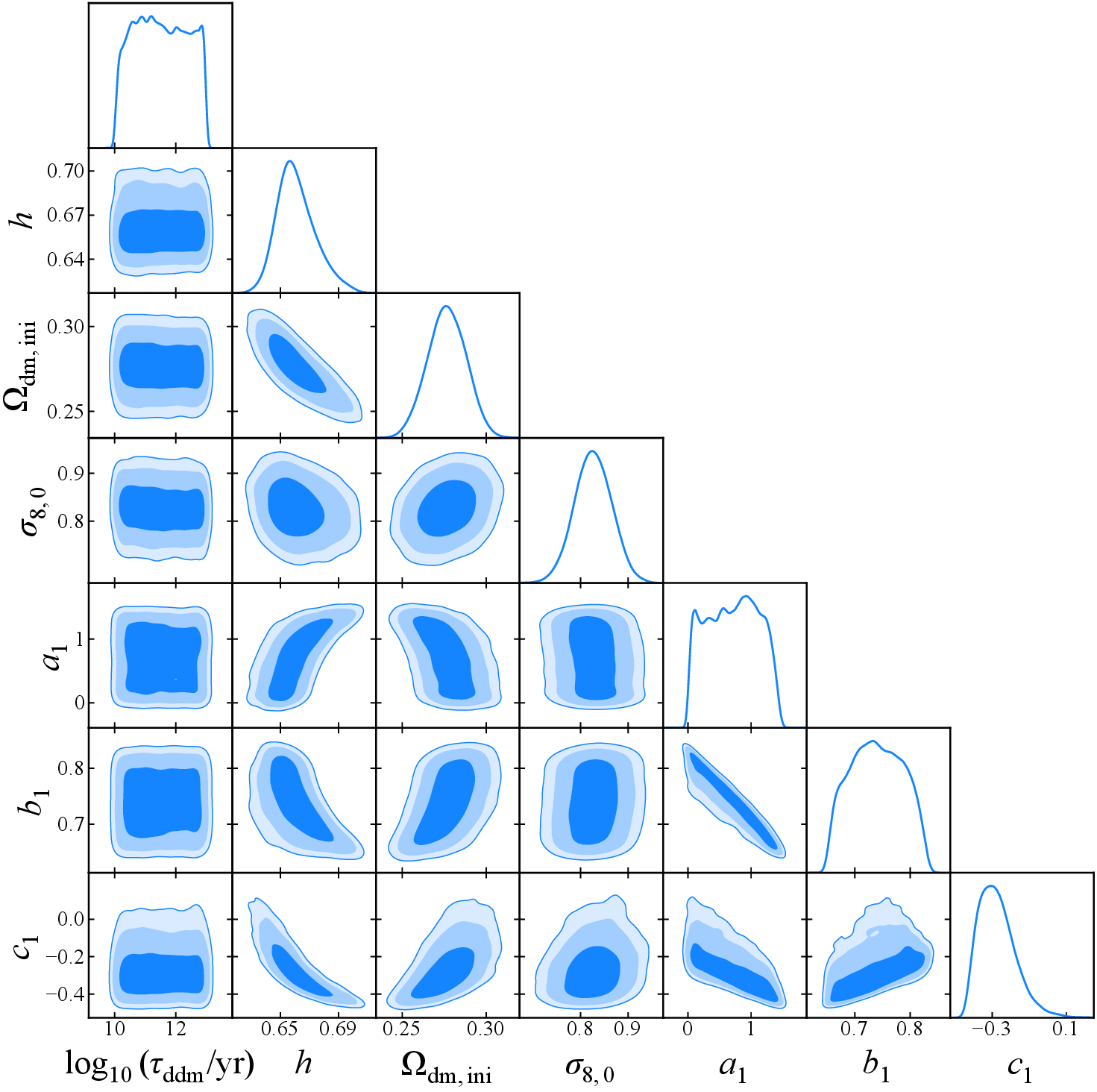}
\includegraphics[width=0.85\columnwidth]{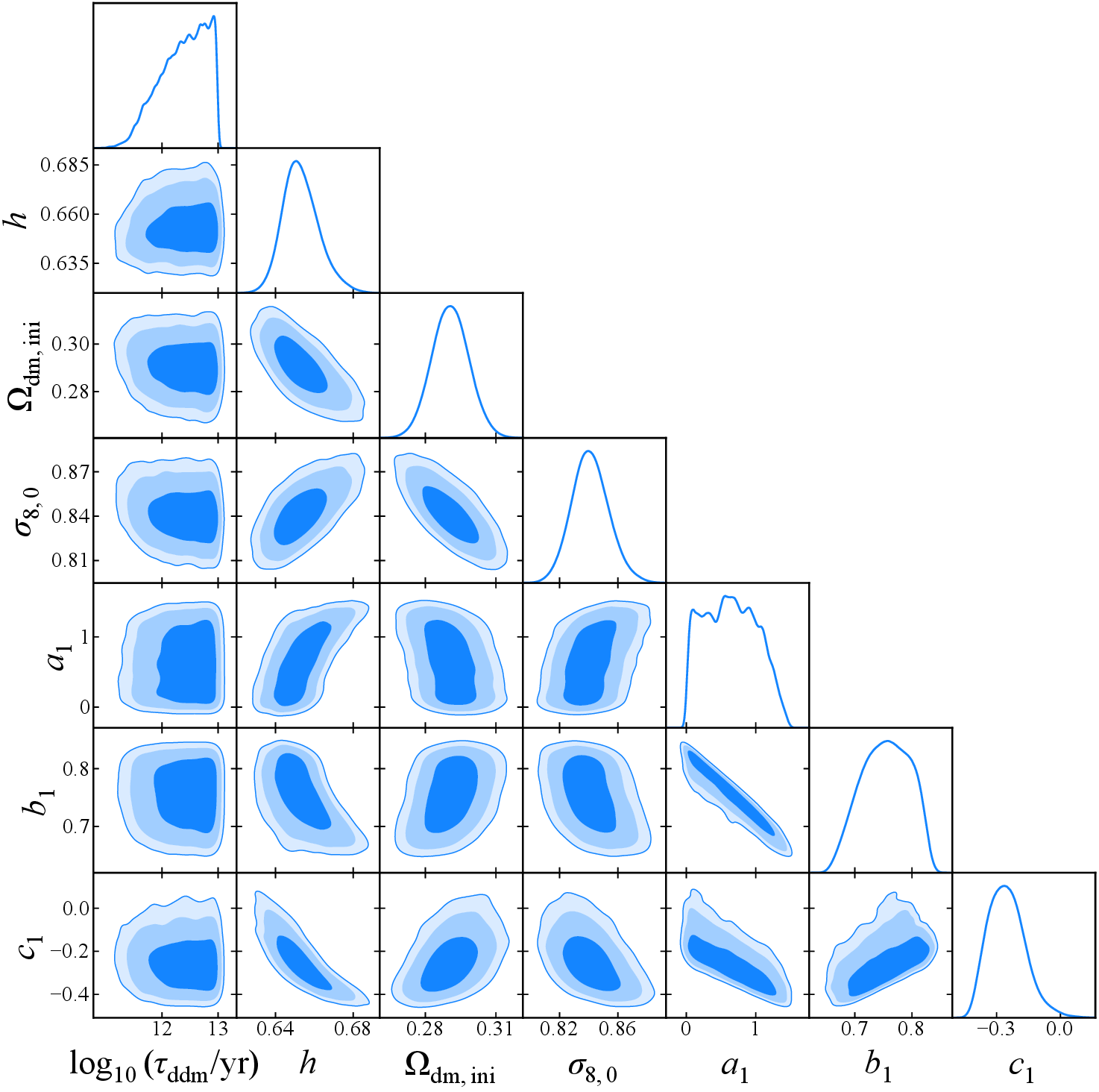}
\captionof{figure}{The figure on the left shows the marginalized DESI DR1 (BAO + shape fit) posteriors for the Pad\'e-DDM model parameters.  Contours show the 1$\sigma$, 2$\sigma$, and 3$\sigma$ credible regions.  The figure on the right shows the joint estimates with HMF data along the BAO + shape fit from DESI DR1. }
\label{fig:corner}
\end{center}
\end{figure*}

\begin{table*}
\begin{ruledtabular}
\begin{tabular}{lcc}
Parameter & DESI DR1 (BAO+ShapeFit) & DESI DR1 (BAO+ShapeFit) with HMF \\
\hline
$\log_{10}(\tauddm/{\rm yr})$ & \posterior{11.53}{0.98}{1.00} & \posterior{12.42}{0.50}{0.40} \\
$h$ & \posterior{0.660}{0.011}{0.014} & \posterior{0.652}{0.008}{0.010} \\
$\Omega_{\rm dm,ini}$ & \posterior{0.276}{0.011}{0.011} & \posterior{0.291}{0.008}{0.008} \\
$\sigma_{8,0}$ & \posterior{0.826}{0.040}{0.040} & \posterior{0.841}{0.012}{0.013} \\
$a_1$ & \posterior{0.73}{0.49}{0.42} & \posterior{0.64}{0.42}{0.41} \\
$b_1$ & \posterior{0.737}{0.050}{0.052} & \posterior{0.756}{0.045}{0.044} \\
$c_1$ & \posterior{-0.287}{0.085}{0.111} & \posterior{-0.260}{0.081}{0.095} \\
\end{tabular}
\end{ruledtabular}
\caption{\label{tab:mcmc_params}Posterior constraints for the decaying-dark-matter plus reconstructed-smooth-dark-energy model. Values are posterior medians with central $68\%$ credible intervals for the DESI DR1 BAO+ShapeFit data set (left)  and for the joint analysis that additionally includes HMF  data from Wang et al. (right). }
\end{table*}
\section{Results}
\label{sec:results}

\begin{figure*}
\includegraphics[width=0.85\columnwidth]{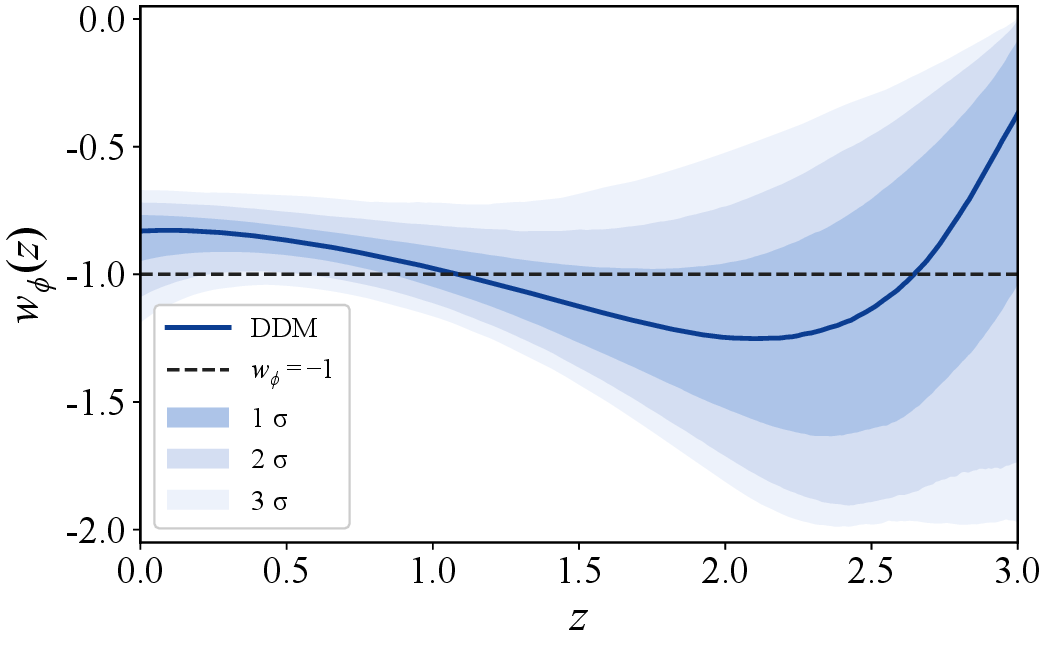}
\includegraphics[width=0.85\columnwidth]{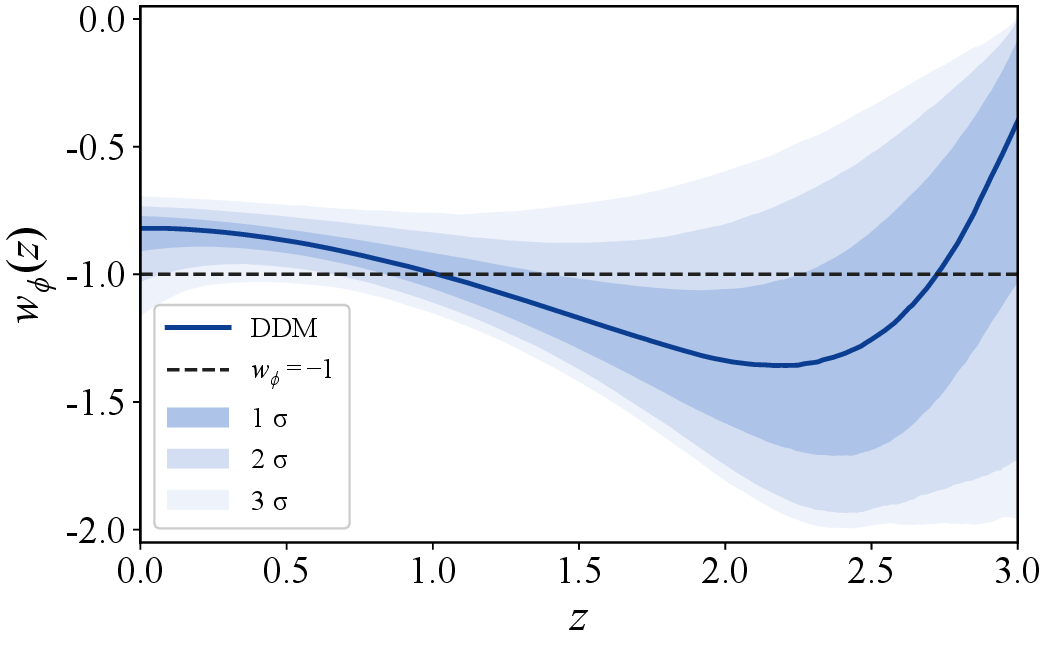}

\captionof{figure}{Posterior reconstruction of the smooth dark-energy equation of state.  Left: DESI DR1 (BAO + shape fit) run.  Right: joint DESI DR1 (BAO + shape fit) with HMF data.  The horizontal dotted reference line is $w_\phi=-1$.}
\label{fig:W_PHI}
\end{figure*}

\begin{figure*}
\includegraphics[width=0.85\columnwidth]{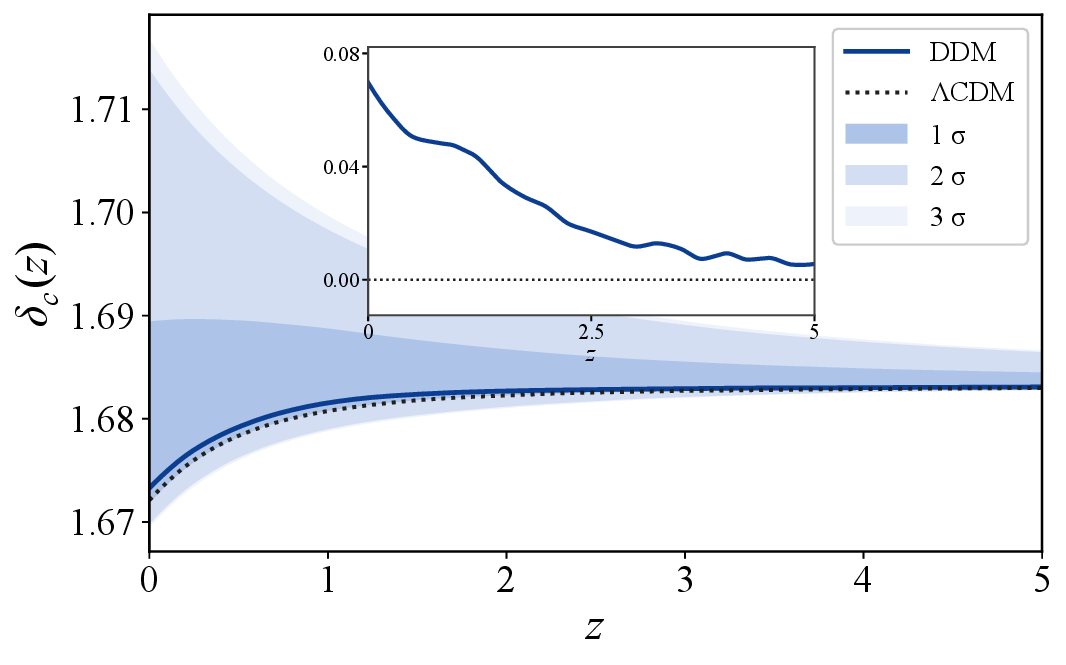}
\includegraphics[width=0.85\columnwidth]{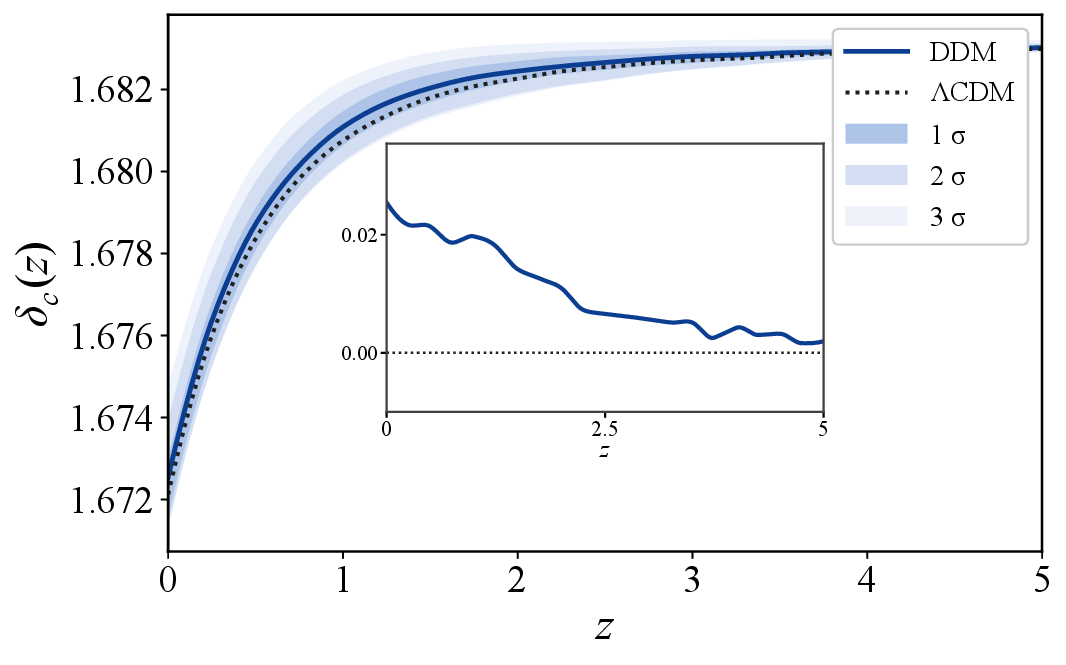}
\captionof{figure}{Reconstructed critical linear collapse threshold $\delta_c(z)$ for the two data combinations considered in this work. Left: DESI DR1 BAO+ShapeFit constraint. Right: joint DESI DR1 BAO+ShapeFit+HMF constraint. The blue curve denotes the DDM reconstruction and the dotted black curve denotes the fitted $\Lambda$CDM reference, with shaded bands showing the posterior credible regions. The inset in each panel shows the corresponding percentage deviation of the DDM prediction from the $\Lambda$CDM case.}
\label{fig:DELTA_C}
\end{figure*}

\begin{figure*}
\includegraphics[width=0.85\columnwidth]{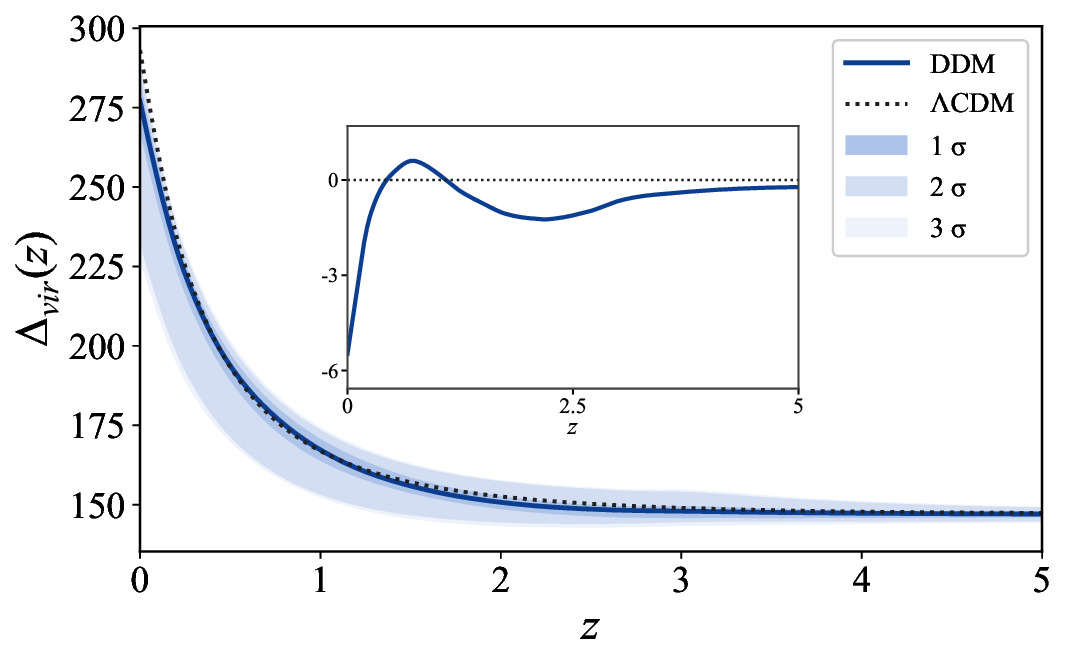}
\includegraphics[width=0.85\columnwidth]{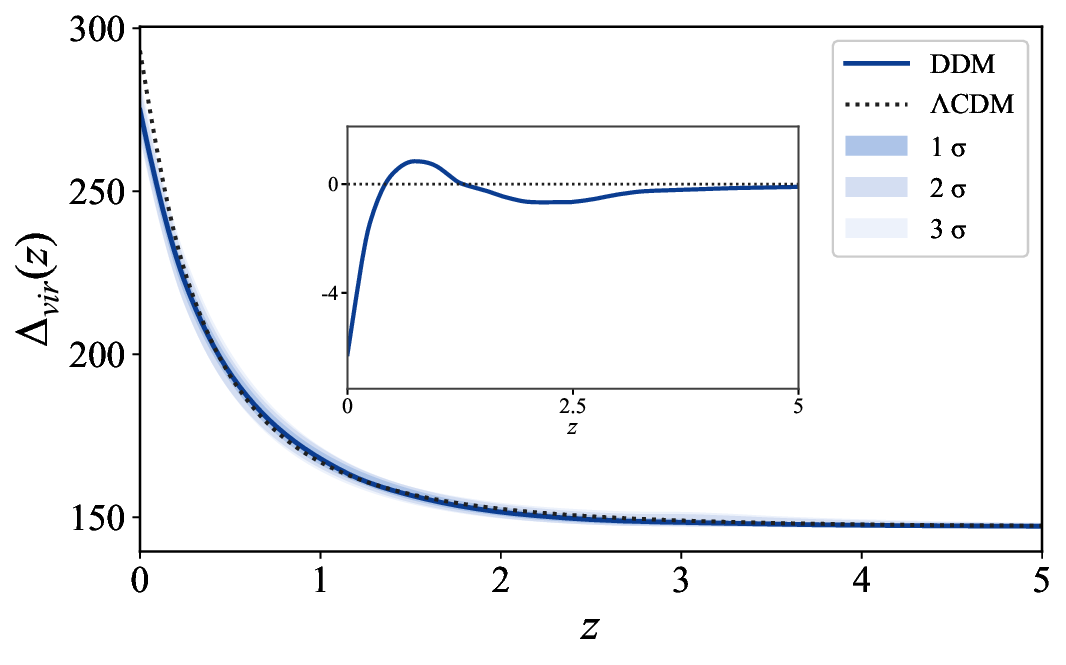}
\captionof{figure}{Reconstructed virial overdensity $\Delta_{vir}(z)$ for the two data combinations considered in this work. Left: DESI DR1 BAO+ShapeFit constraint. Right: joint DESI DR1 BAO+ShapeFit+HMF constraint. The blue curve denotes the DDM reconstruction and the dotted black curve denotes the fitted $\Lambda$CDM reference, with shaded bands showing the posterior credible regions. The inset in each panel shows the corresponding percentage deviation of the DDM prediction from the $\Lambda$CDM case.}
\label{fig:DELTA_V}
\end{figure*}

\begin{figure*}
\includegraphics[width=0.85\columnwidth]{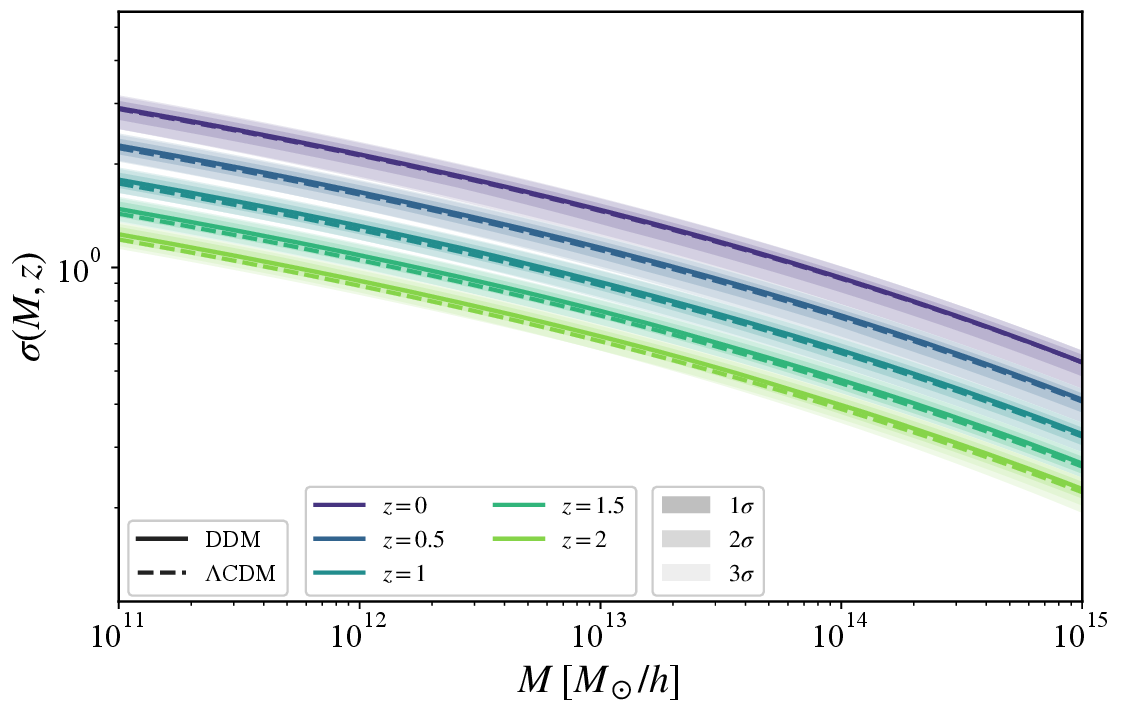}
\includegraphics[width=0.85\columnwidth]{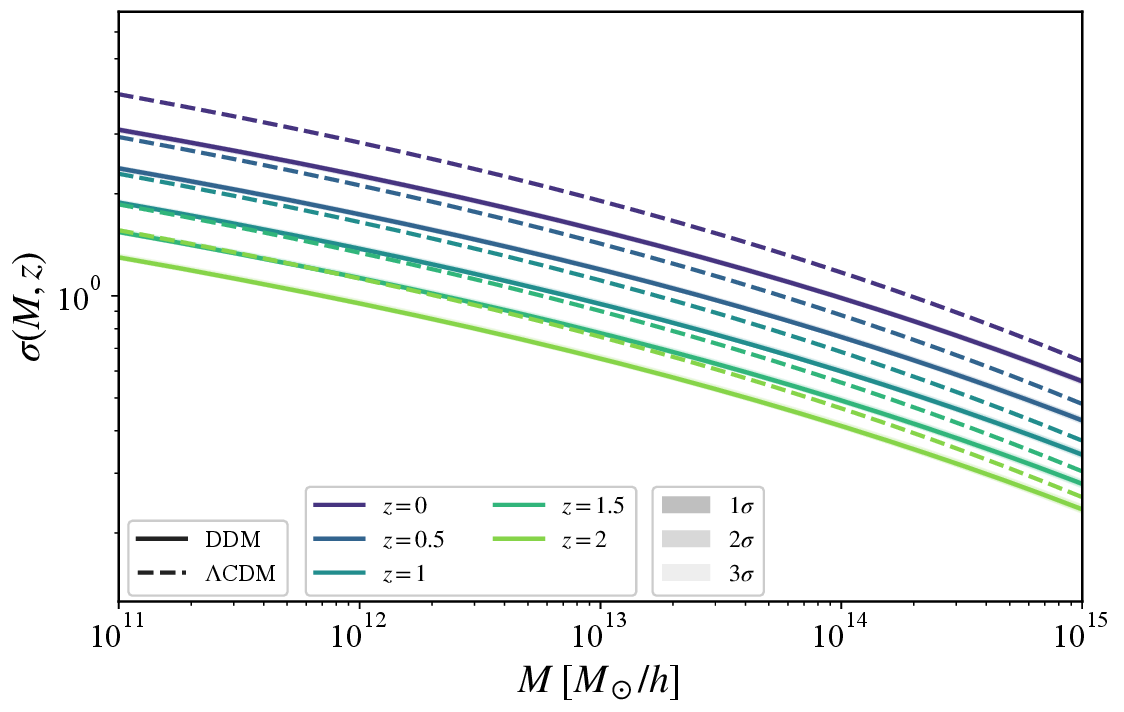}
\captionof{figure}{Mass variance $\sigma(M,z)$ for DDM plus reconstructed dark energy compared with the fitted $\Lambda$CDM model for the same data combination.  Left: DESI DR1 (BAO+ShapeFit).  Right: Joint analysis DESI DR1 (BAO+ShapeFit) with HMF.}
\label{fig:SIGMA}
\end{figure*}

\begin{figure}
\includegraphics[width=0.85\columnwidth]{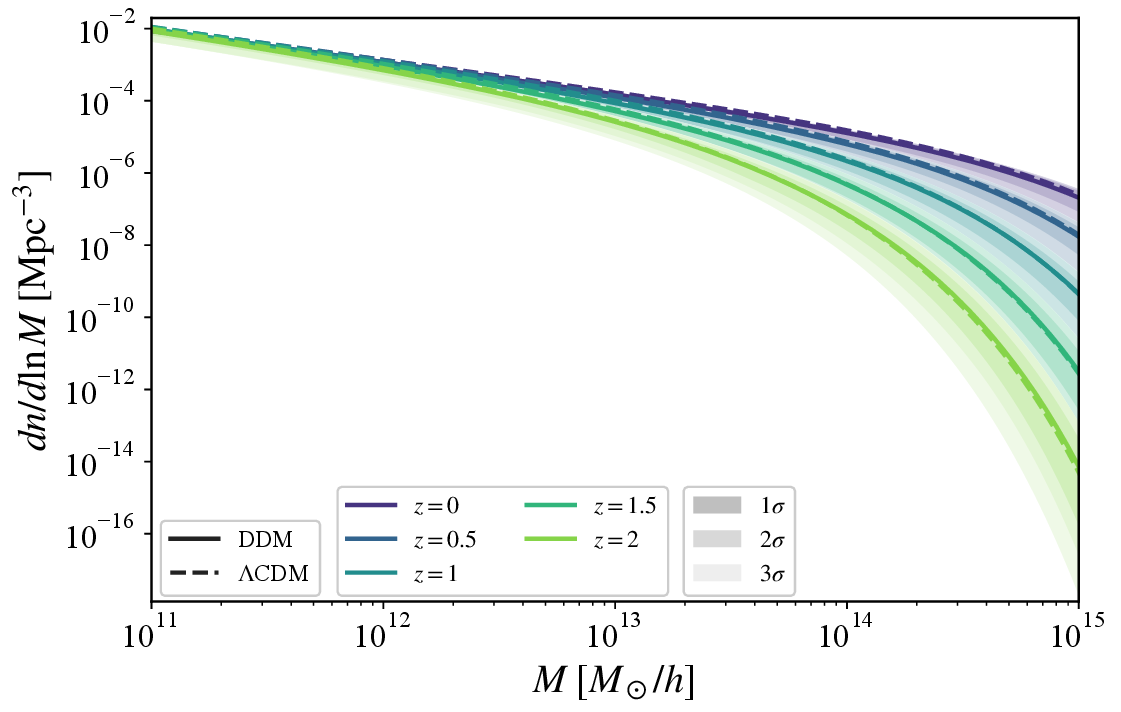}
\includegraphics[width=0.85\columnwidth]{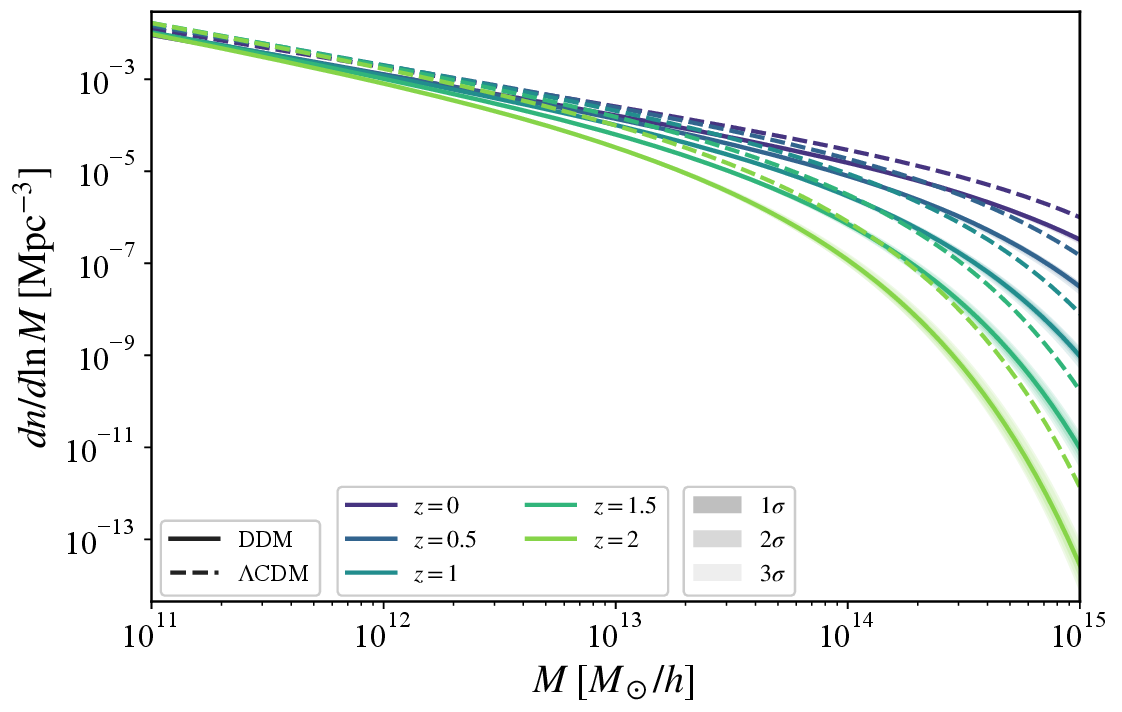}
\captionof{figure}{Differential halo mass function ($\dd n/\dd\ln M$).  Top: DESI DR1 (BAO+ShapeFit) DDM reconstruction compared with the fitted $\Lambda$CDM model.  Bottom: DESI DR1 (BAO+ShapeFit) with HMF reconstruction compared with the fitted $\Lambda$CDM model.}
\label{fig:dndm}
\end{figure}

\label{subsec:posterior}
Figure~\ref{fig:corner} shows the posterior distributions for the parameters $(\log_{10}(\tauddm/{\rm yr}),h,\Omega_{\rm dm,ini},\sigma_{8,0},a_1,b_1,c_1)$. Table~\ref{tab:mcmc_params} shows the corresponding parameter estimates along with $1\sigma$ errors.
Considering only BAO distance and shapeFit data, the decay-lifetime remains broad over the allowed interval $10<\log_{10}(\tauddm/{\rm yr})<13$.  This is expected for a DESI DR1 likelihood with a flexible Pad\'e expansion history since the decay effects are also partially absorbed by the dark energy sector.  The parameters $h$, $\Omega_{\rm dm,ini}$, and $\sigma_{8,0}$ are more localized.  The Pad\'e coefficients show the strongest internal degeneracies.
BAO and ShapeFit constrain the background expansion history and the linear growth of density perturbations, but they are only indirectly sensitive to nonlinear structure formation. 

The halo mass function directly measures the abundance of collapsed dark-matter halos. The inclusion of HMF data substantially improves parameter constraints because halo abundances probe the nonlinear growth of cosmic structures and depend exponentially on the amplitude of matter fluctuations and the collapse threshold. This sensitivity complements the linear information provided by BAO and ShapeFit, breaking degeneracies between dark-matter decay and dark-energy parameters. The constraints in the right column of Table~\ref{tab:mcmc_params} are therefore tighter on both the DDM and dark energy sectors.

\subsection{Reconstructed dark energy equation of state}
\label{subsec:wphi_results}
We ask what kind of dark-energy scenario the data prefer when the matter sector is DDM.
Figure~\ref{fig:W_PHI} 
shows the reconstructed dark-energy equation of state, ($w_\phi(z)$), obtained from the Pad\'e-DDM semi-cosmographic analysis. The median reconstruction exhibits a nontrivial evolution with redshift. At low redshift ($z \lesssim 0.5$ ), the equation of state lies above ($w=-1$), corresponding to a quintessence-like regime. In this regime, the dark-energy component behaves as a canonical scalar field whose kinetic energy remains positive, leading to accelerated expansion without violating standard energy conditions. Unlike a cosmological constant, which has a fixed value ($w=-1$), quintessence allows the dark-energy density to evolve dynamically over time.

At higher redshifts, the reconstructed equation of state crosses the cosmological-constant boundary ($w=-1$) near ($z \simeq 1$) and enters a phantom-like phase, where ($w_\phi<-1$). This transition is commonly referred to as a phantom crossing. Phenomenologically, a phantom equation of state corresponds to an effective dark-energy density that increases with expansion, leading to an expansion history that is more accelerated than that produced by a cosmological constant. In conventional scalar-field models, a stable crossing of the phantom divide is difficult to realize because a canonical field is restricted to ($w \geq -1$). Consequently, evidence for a crossing is often interpreted as indicating either more complex dark-energy physics, effective fluid behavior, modified gravity effects, or parameter degeneracies within the reconstruction.

The median reconstruction remains effectively phantom-like over an intermediate redshift range before crossing back above ($w=-1$) near the upper boundary of the reconstruction redshift window. This second crossing suggests that the preferred expansion history is not described by a simple constant equation of state but instead favors a dynamical evolution that alternates between quintessence-like and phantom-like behavior, such that the equation of state evolves through the phantom divide one or more times during cosmic history.

The uncertainty bands reveal that the reconstruction is significantly better constrained at low redshift than at high redshift. The ($1\sigma$) region remains relatively narrow for ($z \lesssim 1$), indicating that the data provide meaningful constraints on the recent evolution of dark energy. However, the uncertainties increase steadily toward higher redshift, with the ($2\sigma$) and ($3\sigma$) regions eventually occupying a substantial fraction of the allowed prior. This widening reflects the diminishing sensitivity of the DESI data to the detailed evolution of dark energy at earlier epochs, where matter dominates the cosmic energy budget and the influence of dark energy becomes progressively weaker.

While qualitatively showing the same behavior, it is evident here too that inclusion of the HMF data shrinks the uncertainty on the reconstructed dark energy equation of state. 

Overall, the reconstruction from data demonstrates that in a Pad\'e  semi-cosmography with DDM scenario, it does not require the dark-energy component to be exactly consistent with a cosmological constant. At the same time, the broad uncertainty bands at higher redshift indicate that the detailed behavior of $w_\phi(z)$, including the apparent phantom crossings near the edges of the reconstruction interval, should be interpreted with caution. These features are suggestive of dynamical dark-energy behavior but are not yet determined with high statistical significance.

\subsection{Critical density threshold and virial overdensity}
\label{subsec:collapse_results}

Figures~\ref{fig:DELTA_C} and \ref{fig:DELTA_V} gives the spherical-collapse outputs.  The DDM plus reconstructed-DE median $\delta_c(z)$ tracks the $\LCDM$ reference closely over the plotted interval, with only a small sub-percent shift.  The $\LCDM$ curve remains inside the posterior bands, and the redshift dependence is smooth.  This indicates that the critical linear collapse threshold is only weakly modified by the  DDM with dynamical dark-energy backgrounds.

Compared to the critical threshold $\delta_c$, the virial overdensity $\Delta_{vir}$ shows a larger and mildly non-monotonic response.  The DDM median is lower than the $\LCDM$ reference at $z=0$, becomes slightly higher than the reference at intermediate redshift, and then returns toward the reference at high redshift.  This behavior follows the late-time nature of the dark-energy and decay effects. It 
 reflects the competition between dark-matter decay and the reconstructed dark-energy dynamics. At low redshift, the combined reduction in the effective matter density and the enhanced influence of dark energy weaken gravitational collapse, yielding lower virial overdensities than in $\Lambda$CDM. At intermediate redshifts, the modified expansion history temporarily alters the turnaround and virialization process, producing a modest enhancement in $\Delta_{vir}$. At high redshift, dark-energy effects become negligible and only a small fraction of the dark matter has decayed, so the collapse dynamics approach the standard matter-dominated limit and $\Delta_{vir}$ 
 converges toward the $\Lambda$CDM prediction.

The median of the reconstructed $w_{\phi}$  crosses below $-1$  (phantom) approximately around the same redshift where $\Delta_{vir}$  rises above the $\Lambda$CDM counterpart. Thus, the crossover is likely driven primarily by the reconstructed $w(z)$, not by the decaying dark matter itself. Thus, the dynamical dark-energy reconstruction is responsible for the intermediate-redshift enhancement, while DDM mainly suppresses $\Delta_{vir}$ at late times.
The insets in the figure showing the difference from the $\Lambda$CDM demonstrate this.

The posterior bands also become narrower toward high redshift. This is also expected since the Universe becomes increasingly matter-dominated, reducing the influence of both dark-energy evolution and dark-matter decay on the collapse dynamics. As a result, the spherical-collapse solutions approach the nearly universal Einstein–de Sitter limit, making predictions from different posterior samples increasingly similar.

The collapse-sector modifications are relatively small for the posterior considered here. The critical overdensity $\delta_c$ remains close to its $\Lambda$CDM value, while $\Delta_{vir}$ enters the Despali prescription only through the ratio $\frac{\Delta(z)}
{\Delta_{vir}(z)}$ for the adopted spherical-overdensity definition. As a result, these effects contribute only modest corrections to the halo mass function, with the dominant abundance differences arising from changes in the mass variance $\sigma$(M,z) induced by the modified linear matter power spectrum and growth history.

\subsection{Mass variance and Reconstructed HMF}
\label{subsec:sigma_results}

Figure~\ref{fig:SIGMA} shows the reconstructed $\sigma(M,z)$ for the posteriors. We have used the CLASS one-body DDM along with the semi-cosmographic dynamical dark energy everywhere. The $\LCDM$ reference over the mass range is also shown for comparison.  The variance decreases with increasing mass and with increasing redshift.  Relative to $\LCDM$, the DDM posterior median is close to the reference, mildly enhanced over part of the low- and intermediate-mass range, and lower toward the highest masses for the plotted redshift slices.  Since the peak height is $\nu=\delta_c/\sigma$, small mass-dependent changes in $\sigma(M,z)$ can be amplified by the exponential term in Eq.~\eqref{eq:mult}.

Unlike the collapse parameters $\delta_c$ and $\Delta_{vir}$, which vary only weakly across the posterior, the mass variance $\sigma(M,z)$ is more sensitive to the linear growth history and the shape and amplitude of the matter power spectrum. Consequently, the changes in the halo mass function arise through modifications to $\sigma (M,z)$, particularly at the high-mass end. 
For the BAO + ShapeFit posterior, the collapse parameters remain close to their $\Lambda$CDM values, and the corresponding changes in $\sigma (M,z)$  are modest. The inclusion of HMF data, however, directly constrains the abundance of collapsed halos, which depends exponentially on the peak height. Consequently, the joint BAO + ShapeFit + HMF analysis favors larger departures in $\sigma (M,z)$  from the $\Lambda$CDM prediction, while the associated changes in $\delta_c$ and $\Delta_{vir}$
remain comparatively small.

Figure~\ref{fig:dndm} shows the reconstructed HMF for the one-body DDM model with the semi-cosmographic dynamical-dark-energy.  The DDM abundance is close to the $\LCDM$ reference near the lower end of the plotted cluster-mass range and is suppressed in the rare, high-mass tail.  The redshift dependence is most pronounced in the high-mass tail, where rare halos correspond to larger peak heights and are therefore more sensitive to small changes in $\sigma(M,z)$. 

 In the DESI DR1 (BAO+ShapeFit) case, the reconstructed $\sigma(M,z)$ remains relatively close to its fitted $\Lambda$CDM counterpart. In contrast, the joint analysis, which includes halo mass-function data, yields a visibly different $\sigma(M,z)$, particularly at high masses and high redshifts.

 For the DESI DR1 (BAO+ShapeFit) analysis, the HMF differs only moderately from the  $\Lambda$CDM result. At $z=0$, the DDM abundance is lower by about 13--16\% across the plotted mass range. At higher redshifts, both the magnitude and sign of the difference become increasingly mass dependent, reflecting the fact that the fitted $\Lambda$CDM reference is itself optimized to the DESI DR1 (BAO+ShapeFit) data.

The joint analysis with DESI DR1 (BAO+ShapeFit) and HMF  data exhibits a much stronger departure from the  $\Lambda$CDM prediction. The DDM halo mass function lies below the corresponding $\Lambda$CDM result throughout the plotted range, with the suppression increasing systematically with both mass and redshift. At $M=10^{14}\Msun/h$, the median suppression is approximately $>50 \%$  for $z >0.5$, providing the clearest nonlinear signature of the joint analysis.
The largest differences occur at high masses and high redshifts, where the halo abundance is most sensitive to changes in the mass variance.

\section{Discussion}
\label{sec:discussion}

The strong suppression of the high-mass tail of the halo mass function implies that the formation of the most massive gravitationally bound structures is significantly delayed relative to the standard $\Lambda$CDM scenario. Since massive galaxy clusters arise from the rarest peaks in the primordial density field, their abundance is exponentially sensitive to the growth history of matter perturbations. A reduced abundance of such halos therefore reflects less efficient nonlinear structure formation, leading to fewer massive galaxy clusters, a slower assembly of the cosmic web, and a reduced rate of mergers among the most massive halos. This suppression would also be expected to decrease the number of strong cluster gravitational lenses and to lower the cluster contribution to observables such as the thermal and kinetic Sunyaev–Zel'dovich effects \cite{birkinshaw1999sunyaev, carlstrom2002cosmology, refregier2000power}, and X-ray cluster counts \cite{cerardixraycount}. Consequently, measurements of the high-mass end of the halo mass function provide a particularly powerful probe of non-standard dark-sector physics, including decaying dark matter and evolving dark energy.
We also note that the suppression extending to lower halo masses would be expected to delay the formation of the first luminous sources and thereby influence the timing and duration of cosmic reionization.

Several assumptions should be kept in mind.  The posterior uses DESI DR1 BAO and ShapeFit data only, while $r_d$, $\omega_b$, and $n_s$ are fixed.  The dark-energy component is treated as smooth and unclustered, and the virialization prescription is analytic.  In the HMF calculation, the CLASS power spectrum includes the semi-cosmographic dynamical dark-energy background. 

We also note that a limitation of the present analysis is that the published halo mass-function measurements \cite{DR9} do not provide a covariance matrix. We therefore adopt a diagonal covariance matrix constructed from the reported uncertainties. This neglects correlations between different halo-mass bins and may affect the estimated parameter uncertainties, although the qualitative trends and best-fit predictions are expected to remain robust. Future analyses using the full covariance matrix, when available, will enable a more reliable quantification of the statistical significance of the inferred constraints.

\section{Conclusions}
\label{sec:conclusion}

We have propagated posteriors for a one-body DDM plus semi-cosmographic dark-energy model obtained from two data combinations: (1) DESI DR1 BAO+ShapeFit and (2) DESI DR1 BAO+ShapeFit+HMF. We use these posteriors to constrain DDM parameters and to reconstruct the dark-energy EoS, spherical-collapse observables, and halo mass function. The main conclusions are:

\begin{enumerate}
\item The DESI DR1 posterior leaves the DDM lifetime weakly localized because a flexible Pad\'e background can absorb part of the smooth expansion effect of decay. Inclusion of the HMF data improves the constraints significantly.

\item The residual dark-energy reconstruction can differ from $w_\phi=-1$, but the high-redshift evolution of dark energy remains weakly constrained.

\item The critical collapse threshold $\delta_c(z)$ remains close to the reference $\LCDM$ value, so its direct effect on the HMF is small.
\item The virial overdensity has a clearer, mildly non-monotonic response and approaches the reference behavior at high redshift.  Inclusion of the HMF data in the posterior reduces the uncertainties and makes the departure from the $\Lambda$CDM behaviour more evident.

\item A significant change in the halo-abundance response is seen in the joint analysis, especially at the large mass tail. 
\end{enumerate}

We conclude by noting that our results demonstrate that halo abundances provide a sensitive nonlinear probe of coupled dark-sector physics, substantially tightening constraints beyond those obtained from geometric observables alone. This framework establishes a direct connection between reconstructed dark-energy dynamics, decaying dark matter, and the formation of cosmic structure, and highlights the potential of upcoming large-scale structure surveys to test non-standard cosmological models involving decaying dark matter and dynamical dark energy.

\section {Data Availability}
The DESI DR1 BAO and ShapeFit compressed data used in this work are publicly available from the DESI Collaboration.

\begin{acknowledgments}
T.G.S acknowledges financial support from the ARG–MATRICS project ANRF/ARGM/2025/002607/TS.

M.Y. acknowledges financial and computational support from Birla Institute of Technology and Science, Pilani.
\end{acknowledgments}

\bibliographystyle{apsrev4-2}
\bibliography{references}

\end{document}